\documentclass[onecolumn,usenatbib,useAMS]{mnras}

\usepackage{graphicx}
\usepackage{amsmath}
\usepackage{amsfonts}
\usepackage{amssymb}%
\providecommand{\U}[1]{\protect\rule{.1in}{.1in}}

\usepackage[yyyymmdd,hhmmss]{datetime}

\usepackage{lineno}

\newcommand{\beq}{\begin{equation}}
\newcommand{\eeq}{\end{equation}}
\newcommand{\ba}{\begin{array}}
\newcommand{\ea}{\end{array}}

\newcommand{\ee}{\epsilon_{e,0}}

\def\be{\begin{equation}}
\def\ee{\end{equation}}

\def\gsim{\raisebox{-0.3ex}{\mbox{$\stackrel{>}{_\sim} \,$}}}

\newcommand{\pder}[2][]{\frac{\partial}{\partial#2} \left (#1 \right )}

\usepackage{color}

\title[Radiative magnetic reconnection]{Radiative striped wind model for gamma-ray bursts}
\author[B\'egu\'e, Pe'er \& Lyubarski]{D. B\'egu\'e$^{1,2}$, A. Pe'er$^{3}$ \& Y. Lyubarski$^{4}$
\\
$^1$The Oskar Klein Centre for Cosmoparticle Physics, AlbaNova, SE-106 91 Stockholm, Sweden \\
$^2$Department of Physics, KTH Royal Institute of Technology, AlbaNova University Center, SE-106 91 Stockholm, Sweden \\
$^3$Physics Department, University College Cork, Cork, Ireland \\
$^4$Physics Department, Ben-Gurion University, P.O.B 653, Beer-Sheva 84105, Israel}

\begin{document}
\maketitle
\begin{abstract}
In this paper we revisit the striped
 wind model in which the wind is accelerated by magnetic
reconnection. In our treatment, radiation is included as an independent
component, and two scenarios are considered. In the first one,
radiation cannot stream efficiently through the reconnection layer,
while the second scenario assumes that radiation is homogeneous in the
striped wind. We show how these two assumptions affect the
dynamics. In particular, we find that the asymptotic radial evolution
of the Lorentz factor is not strongly modified whether radiation can
stream through the reconnection layer or not. On the other hand, we show that the width, density and temperature of the reconnection layer are strongly dependent on these assumptions.
We then apply the model to the gamma-ray burst context and
find that photons cannot diffuse efficiently through the reconnection layer below radius $r_{\rm D}^{\Delta} \sim 10^{10.5}$ cm, which is about an order of magnitude below the photospheric radius.
Above $r_{\rm D}^{\Delta}$, the dynamics asymptotes to the solution of the scenario in which radiation can stream through the reconnection layer. As a result, the density of the current sheet increases sharply, providing efficient photon production by the Bremsstrahlung process which could have profound influence on the emerging spectrum. This effect might provide a solution to the soft photon problem in GRBs.  

\end{abstract}


\section{Introduction}

The solution to the compactness problem of gamma-ray bursts (GRBs)
requires the emitting plasma to expand at ultra-relativistic speed,
with a Lorentz factor $\Gamma \geq 100$ (for reviews see \textit{e.g.}
\cite{Pir99,Mes06}, and more recently \cite{KZ15,Pee15}). The
acceleration can be provided either by the thermal pressure if
magnetic fields are sub-dominant, $U_{\rm B} \ll U_{\rm \gamma}$
\citep[this is the classical ``fireball'' model;
  see][]{Pac86,Pac90,RM92,RM94,PSN93}, or at the expence of magnetic
energy, if $U_{\rm B} \gg U_{\rm \gamma}$
\citep{SDD01,Lyu06,NMF07,TMN08,KVK09,MU12}. In addition, the high
efficiency of GRB prompt phase \citep{CFH11} implies that the plasma
should be an efficient emitter during and/or after its acceleration.
This requires the plasma energy to be dissipated, either via shocks
\citep{RM92,RM94,DM98} or in magnetic reconnection
\citep{Uso92,Tho94,DS02}.

The question of magnetization in GRB outflows is still open. The
initially suggested non-magnetized models face several
difficulties. These include: (1) the low efficiency energy output by
the central engine through neutrino annihilation \citep{DPN02,SLG15};
(2) the low efficiency of kinetic energy dissipation by shocks
\citep{KPS97,PSM99}; and (3) if the energy is dissipated in shocks, the difficulty of the synchrotron
process to account for steep and narrow spectra \citep{GCL00,AB15}.
On the contrary, Poynting-flux models do not suffer from these
problems. In particular, kinetic energy does not need to be
dissipated, provided that magnetic energy can be converted to
radiation, for instance through magnetic reconnection. In addition,
the rotational energy of a black-hole can be efficiently tapped by the
Blandford-Znajeck mechanisms \citep{BZ77}, resulting in a Poynting
flux dominated outflow. This mechanism is thought to play a central
role in the physics of active galactic nuclei (AGN)
\citep{BBR84,WC95}, and accreting systems such as X-ray emitting
binaries (XRBs) \citep{NMF07,TMN08,KVK09,TNM10,MU12}.

There are two different mechanisms which could operate in accelerating
Poynting-flux jets to relativistic velocities, depending on the
topology of the magnetic field lines. First, the acceleration can be
powered by the expansion of the magnetic field lines, provided that particles are
attached to the field lines. This expansion follows the amplification of the
magnetic field and twist of the lines by the rotation of the central black-hole
\citep{TMN08,TNM10}. Second, the acceleration of GRB jets might be
powered by reconnection of the magnetic field lines
\citep{Tho94,LK01,SDD01,DS02}, similar to pulsar wind nebulae
\citep{Cor90}. A fast rotating neutron star having rotational axis
misaligned with its dipolar moment naturally produces a
\textit{striped wind} above the light cylinder, consisting of cold
regions with alternating magnetic field separated by hot current
sheets. Furthermore, striped winds may also be produced by black holes
\citep{SDD01}. Such a wind is accelerated by the reconnection of
magnetic field lines with opposite polarity. Early models of the
\textit{striped wind} were developed for pulsars and more specifically
for the Crab nebula \citep{Mic71, Cor90,Mic94,LK01}, and scaled to GRB
physics by \cite{DS02,Dre02, Gia06} and \cite{Gia12}.

Although the dynamics of the \textit{striped wind} model have been
studied by several authors over the years, possible effects of
radiation have not been addressed yet. This could be of particular
importance in the context of GRBs, since a GRB central engine is
expected to be heavily loaded with baryons, implying that the
\textit{striped wind} is initially optically thick, as opposed to the
pulsars case. Therefore, the photons emitted inside the hot current sheet
(or reconnection layer) are coupled to the flow until their escape at
the photosphere. The properties of these photons will therefore be
manifested as they emerge from the photosphere
\citep{Goo86,Pac86,ANP91}.

An observed photospheric signal within the framework of the
\textit{striped wind} model is expected to differ from signals
expected within the framework of existing photospheric models, due to
the fact that this model naturally possesses a non-homogeneous density
profile. It may therefore suggest a natural solution to reconcile
efficient photon production in the dense current sheets together with
ultra-relativistic motion.  Indeed, photopheric models rely on
dissipation of energy below the photosphere to account for the soft
low-energy spectral slopes and the high efficiency of the prompt phase
\citep{CFH11}. However, the observed spectral peak at the energy of only a few hundred keV implies efficient photon production at intermediate distances $10^{9} - 10^{10}$ cm, and the Lorentz factor of the outflow should not exceed $\sim$10 in this zone \mbox{\citep{Bel13,VLP13}}.

Furthermore, considering radiation as an independent fluid component
allows to study its effects on the dynamics, on the acceleration rate,
and on the evolution of the wind internal structure
\citep{Cor90,LK01}.  In the context of GRBs, this had never been done
before. Previous works by \citet{SDD01}, \citet{DS02}, and
\citet{Dre02} studied the dynamics of the plasma in the
\textit{striped wind} model, neglecting its internal
structure. \citet{DS02} included radiative losses in their numerical
treatment, assuming that (1) the internal energy is uniform in the
striped wind, and (2) the emissivity is constant, equal to an
arbitrary value.  The goal of this paper is to bridge this gap by
studying how the dynamics and the internal \textit{striped wind}
structure are modified by the existence of a strongly coupled
radiation field, as is expected below the photosphere.

In the paper, we study two limiting cases for the distribution of
radiation in the \textit{striped wind}. Radiation can either (1) be confined to the current sheet (hereafter case I) or (2) fill
the full outflow   (hereafter case II). We show that
acceleration proceeds at a similar rate in both cases. In particular,
the Lorentz factor increases proportionally to the radius as $\Gamma
\propto r^{1/3}$, as was first found by \cite{DS02}. In addition, we find that the proportionality
coefficient is weakly dependent on the radiation distribution, with
only $\sim 5\%$ difference between the two scenarios, independent of
any parameters characterizing GRB outflows. On the other hand, we show
that the internal structure of the wind is very different in these
limiting cases. The current sheets width and density are respectively
much thinner and higher when radiation is assumed to stream through
the current sheets into the magnetized (cold) region, because in this case, the magnetic pressure is balanced only by the gas pressure in the sheet whereas the gas temperature remains relatively low being in equilibrium with the radiation. We compute the rate of photon diffusion and find that the heat produced
by magnetic reconnection remains confined in the current sheet up to
the radius $r_{ \rm D} \sim 10^{10.5} \rm ~cm$, smaller than the
photospheric radius. Above $r_{ \rm D}$, the heat can be transported
and distributed in the magnetized region. Therefore the flow is in the regime I below $r_{ \rm D}$ and in the regime II above this radius. We show that due to the high plasma density in the current sheet in the regime II, the radiation is efficiently thermalized via the Bremsstrahlung emission/absorption, which could have profound effect on the emergent spectrum.

The paper is organised as follow. In section 2, a description of the
\textit{striped wind} with its governing equations is given. Then,
section 3 presents the asymptotic solution to the governing
equations. Section 4 deals with the rates of photon diffusion and of
photon production mechanisms. The implications of our work on the
physics of GRBs are discussed in section 5, and the conclusion
follows. As the complete derivation of the equations is cumbersome and
not required to obtain a clear physical picture, it is presented in
the appendix.

\section{Basic properties of the model}

\subsection{Parameters of the striped wind}

In the collapsar scenario \citep{Woo93,WH06}, a magnetar or a rotating
black-hole is naturally expected to form in the center of a GRB
progenitor. It will produce a \textit{striped wind}
\citep{Tho94,SDD01}, provided that the central region is highly
magnetized.\footnote{If the central object is a neutron star, it is
  necessary that the obliquity between the rotational axis and the
  dipole moment of the compact object is non-null.} Below the
light-cylinder, defined by its radius $r_L = c / \Omega$, with $c$ the
speed of light and $\Omega$ the angular velocity of the neutron star,
the field lines are closed and the plasma is in co-rotation with the
neutron star \citep{Mic69}. However, above the light-cylinder, the
field lines must be open \citep{GJ69}. In the aligned rotator case,
current sheets are formed in the dipolar equatorial plane (which is
also the spin equatorial plane) above $r_L$, as this plane separates regions
with different magnetic polarity. Indeed, the open field lines on each
side of the equatorial plane are attached to a different pole of the
central neutron star.

In the case of an oblique rotator, the current sheet is not steady. At
the spin equator, the polarity of the magnetic field alternates
\citep{Mic71,Cor90,Spi06}. Therefore, above the light-cylinder the
wind can be described by regions with nearly toroidal magnetic field
of alternating polarity, separated by thin non-magnetized current
sheets \citep{Mic71}. The length-scale $l_0$ on which the polarity
alternates is comparable to the rotation period of the neutron star
times the speed of the plasma above the light-cylinder, which is
nearly $c$, therefore $l_0 \sim 2\pi c/\Omega =2\pi r_L$.
The relative strength of the toroidal magnetic
field of alternating polarity depends on the obliquity between the
spin axis and the magnetic moment. The strength is roughly equal
for an obliquity of 90$^{\circ}$, the only situation considered in
this paper.\footnote{The case of a non-orthogonal rotator was
  partially addressed for pulsars by \cite{Cor90}.}

\begin{figure}[htb]
\centering
\resizebox{0.9\linewidth}{!}{\input{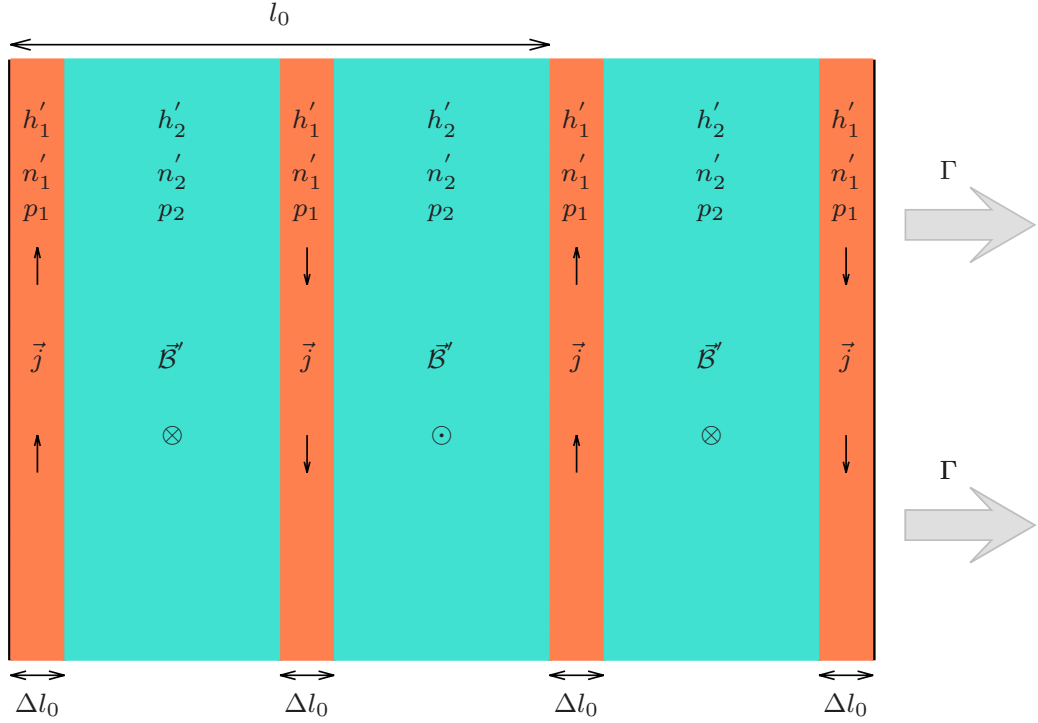}}
\caption{Details of the small length-scale structure of the
  \textit{striped wind}: magnetized regions (in blue) of alternating
  polarity are separated by dense current sheets (in red). Both the
  pressure and the entropy have contributions from the gas and from the
  radiation: $p_i = p_{i,\rm g} + p_{i, \rm rad}$ and $h_i^{'} =
  h_{i,\rm g}^{'} + h_{i, \rm rad}^{'}$. }
\label{fig:cartoon_striped}
\end{figure}

As a result of the current sheet oscillations at the spin equator,
the outflow is composed of two regions, as shown
in Figure \ref{fig:cartoon_striped}. The first is the {\it current
  sheet}, in which the reconnection of the magnetic field lines takes
place. This region is non magnetized\footnote{The assumption that the current sheet is not magnetized at all is an idealization. In reality, the magnetic field smoothly changes sign within the sheet. As it is shown by \citet[see the very end of the Appendix]{LK01}, the striped wind in the general case  is described by the same equations as the idealized wind with non-magnetized current sheets with appropriately normalized parameters.}, hot, and is characterized by
high comoving density, denoted here by $n_1^{'}$ (here and below, $X'$
represent quantities measured in the comoving frame). Being hot, the
pressure in this region is dominated by its thermal component.  We
normalize the width of the current sheet, $l_{\rm cs}$ relative to the
alternating field width $l_0$ by writing $\Delta = l_{\rm
  cs}/l_0$. Typically, $\Delta \ll 1$, which can intuitively be
understood as this region is compressed by the magnetic pressure of
its neighbouring regions and is therefore much narrower than the
magnetized regions. The second region is the magnetized region, which
is characterized by comoving density $n_2'$. This density is lower than the
comoving density in the current sheet, $n_2^{'} < n_1^{'}$.  The
pressure in this region is dominated by its magnetic component,
${\mathcal{B}^{'}}^2/(8\pi)$.\footnote{In this paper, all quantities
  measured in the current sheet are labelled with the sub-script 1,
  while the quantities measured in the magnetized regions are labelled
  2.}

Photons are emitted within the hot, dense reconnection layer.  Up to
the photospheric radius, which is typically at $\sim 10^{11.5}$~cm (see
Section 4.1 below), GRB outflows are optically thick. However, due to
the narrowness of the reconnection layer, it is possible that
photons diffuse into the magnetized region at radii much smaller than
that. These photons thus transfer energy and entropy from the hot
current sheet and redistribute them inside the colder magnetized region.
The diffusion of photons between the different layers below the
photosphere may therefore have a significant impact on the dynamics
and on the internal wind structure. In order to study these energy and
entropy transfers, we consider the radiation field as an explicit
independent fluid component.

At radii much smaller than the photospheric radius, as considered in
this work, the photon and particle fields are strongly coupled. We
thus assume instantaneous redistribution of heat and entropy between
photons and particles in the different regimes. This implies that the
photon and gas in each regime assume similar temperature (in the
general case, clearly $T_1' \neq T_2'$).

In describing the dynamics, we consider efficient acceleration due to
magnetic reconnection that begins at radius $r_0 \gsim r_L$. As a
boundary condition, we assumed that at $r_0$ the current sheet width
is very small, $\Delta_0 \sim 0$.\footnote{Note that there is no
  striped wind below the light cylinder, as the magnetic field is in
  co-rotation with the central compact object.} The outflow is
accelerated at the expense of the magnetic energy, which is dissipated
by reconnection in the current sheets. As the magnetic pressure drops,
the current sheet width increases, eventually filling the full outflow
when all the magnetic energy was dissipated and converted to heat
and bulk velocity.

Close to the light-cylinder, the structure of the magnetic field is
not well-known. However, the flow is stretched in the transverse direction so that far
beyond the light cylinder, the field becomes predominantly toroidal and the current sheets
are perpendicular to the flow velocity. If the field is not dissipated, it varies as
$B\propto \varpi^{-1}$, where $\varpi$ is the  cylindrical radius of the flow (the striped wind forms by oscillations of the current sheet at the spin equator).
Assuming for simplicity that the flow is radial, one can write
\begin{linenomath*}
\begin{equation}
\mathcal{B}(r) = \mathcal{B}_0 \frac{r_0}{r},
\end{equation}
\end{linenomath*}
where $\mathcal{B}_0$ is the magnetic field at $r_0$.\footnote{ The
  procedure developed in this paper is accurate for radii much larger
  than the light-cylinder. Yet, we normalize all quantities to their
  value at $r= r_0$, which is of the same order as $ r_{\rm L}$.}  At
the light-cylinder, the luminosity $L$ (per steradians) is assumed to
be dominated by its Poynting flux component,
\begin{linenomath*}
\begin{equation}
L \approx L_{\rm pf} = \frac{ c r_0^2 \mathcal{B}_0^2}{4\pi}.
\end{equation}
\end{linenomath*}
The initial magnetization of the outflow, $\sigma_0$ is defined as the ratio
of the Poynting flux luminosity (which is equal to the total luminosity at $r_0$),
to the kinetic-energy flux at $r_0$
\begin{linenomath*}
\begin{equation}
\sigma_0 = \frac{L}{ m_p c^3 n_0^{'} r_0^2 \Gamma_0^2} =
\frac{\mathcal{B}_0^2}{4\pi m_p c^2 n_0^{'} \Gamma_0^2 },
\end{equation}
\end{linenomath*}
where $\Gamma_0$ and $n_0^{'}$ are the Lorentz factor of the outflow
and the comoving density at $r_0$, and $m_p$ is the mass of a proton.

\subsection{Governing equations of the \textit{striped wind}}

The \textit{striped wind} can be described by two separate
length-scales. First, the nearly periodicity of the magnetic pattern
defines the short length-scale, comparable to $l_0$. It describes the
polarity change of the magnetic field and the current sheets, which
consist of the internal structure of the \textit{striped
  wind}. Second, the reconnection sheet growth (corresponding to the
characteristic magnetic field decay length) defines a second
length-scale 
\begin{linenomath*}
\begin{align}
R = r/r_L \gg 1,
\end{align}
\end{linenomath*}
much larger than the first one. This
second length scale characterizes the plasma expansion.

The hydrodynamical quantities of both the current sheet and the
magnetized region do not vary significantly over the short
length-scale, corresponding to one period of the \textit{striped wind}
internal structure.  In order to obtain the set of equations
describing the parameter evolution on the large length-scale $R$, we
therefore average the conservation equations over the short
length-scale. Our treatment is therefore analogue to treatment of one
full period of the \textit{striped wind} as a single fluid
element (with internal structure). While we provide here the basic set of equations, we give
the full details of the derivation of the relativistic radiation MHD
equations in Appendix A, while the two scales expansion used to obtain
the equations solved below is fully described in Appendix B.
Application to the \textit{striped wind} is given in Appendix C.

We assume that ideal MHD conditions hold in the magnetized region.
The flux freezing condition is a direct consequence of the Faraday
equation and of the continuity equation (see the derivation of
Equation \ref{eq:appendix_fluxfreezing_derivation} in Appendix B). On
the large length-scale $R$, it reads:
\begin{linenomath*}
\begin{align}
\pder[\frac{\mathcal{B}^{'}}{R n_2^{'}}]{R} = 0, \label{eq:fluxfreezingderivative}
\end{align}
\end{linenomath*}
where $\mathcal{B}^{'} =\mathcal{B}/\Gamma$ is the comoving magnetic
field, which decays only over the slow time-scale. 

The averaged over the striped period continuity equation becomes 
\begin{linenomath*}
\begin{align}
\pder[\Gamma \beta R^2 (\Delta n_1^{'}+(1-\Delta)n_2^{'})]{R} = 0, 
\label{eq:continuityderivative}
\end{align}
\end{linenomath*}
where $\beta = (1-\Gamma^{-2})^{1/2}$ is the normalized outflow speed.
The averaged energy-flux equation reads
\begin{align}
\pder[\beta \Gamma^2 R^2 \left ( h_1^{'} \Delta
  +(1-\Delta)\left(h_2^{'}+\frac{{\mathcal{B}^{'}}^{2}}{4\pi} \right )
  \right )]{R} = 0, \label{eq:energyderivative}
\end{align}
where $h_i^{'} = \epsilon^{'}_{i} + p_i$ is the total enthalpy, including the contribution from
the gas and from the radiation $h_i^{'} = h_{i,\rm g}^{'} + h_{i, \rm
  rad}^{'}$.  Here, $\epsilon_i^{'}=\epsilon^{'}_{i, \rm rad} + \epsilon^{'}_{i, \rm g}$ is
the internal energy density, including both the internal energy
density of the radiation and of the gas. The entropy equation is given by (angular brackets mean averaging over le stripe wind)
\begin{align}
\frac{\langle h^{'} \rangle}{R^2} \pder[\beta \Gamma R^2]{R} + \beta \Gamma \pder[\langle \epsilon^{'} \rangle ]{R} + \frac{1}{4\pi R} \big \langle \mathcal{B}^{'} \pder[\beta \Gamma R \mathcal{B}^{'}]{R} \big \rangle =0. \label{eq:entropyequation}
\end{align}
The set of magneto-hydrodynamical equations is completed by the perfect gas law
$p_{i,\rm g} = k_{\rm B} T_i^{'} n_i^{'}$ where $k_{\rm B}$ is the
Boltzmann constant, by the relativistic equation of state for
radiation $\epsilon_{i, \rm rad}^{'} = 3 p_{\rm rad}$ and by an
equation of state for the gas
\begin{align}
p_{i,\rm g} = (\hat \gamma -1)(\epsilon^{'}_{i, \rm g} - n^{'}_i m c^2),
\end{align}
where $\hat \gamma$ is the adiabatic index. The temperature in the flow is non-relativistic, as can be checked a posteriori (maybe with the exception of very small radii which are of no interest here). Therefore we could use $\hat \gamma$=5/3. The index $i=1,2$ describes the two regions (current sheet and magnetized regions).

Finally, a prescription for the magnetic dissipation is required to close the system. In
\cite{Cor90} and \cite{LK01}, the reconnection layer width is set to
be two times the Larmor radius of an electron in the reconnection
layer. Instead in this paper, we chose to assume that the reconnection
rate is constant with the radius and is equal to a fraction $\epsilon$ of
the Alfv\'en velocity\footnote{This fraction of the Alfv\'en velocity
  ($\epsilon$) is not to be confused with the energy density of the
  gas $\epsilon_{i, \rm g}^{'}$. Additionally, we note that $\epsilon$
  only appears associated to the angular velocity of the central
  objects $\Omega$, to form the parameter $(\epsilon \Omega)$.},
roughly equal to $c$ in highly magnetized outflows. Following
\cite{Dre02} and \cite{DS02}, the reconnection rate is parametrized as
\begin{align}
\pder[\beta \Gamma R \mathcal{B}^{'}]{R} = -A \frac{R \mathcal{B}^{'}}{\Gamma} \label{eq:reconnectionrate}
\end{align}
where $A = (\epsilon \Omega) r_{\rm L} /(2\pi c) = \epsilon/(2\pi)$ is a constant\footnote{The multiplication by $r_{\rm L}$ is present to set $A$ in proper units of normalized radius $R$.}.
We further motivate this choice and discuss its impact on our results in the discussion section.

\section{Solution to the conservation equations during the accelerating phase.}
\label{sec:approxsol}

The full solutions to the above set of equations could be obtained only numerically (see sect. 3.3). However, one can find analytical asymptotics for the intermediate zone, where the flow is already significantly accelerated, $\Gamma\gg\Gamma_0$ but still remains Pointing dominated, $\sigma\equiv B^2/4\pi m_p n^{'} c^2\Gamma^2\gg 1$. The last condition implies $\Delta\ll 1$.  
Equations \ref{eq:continuityderivative} - \ref{eq:reconnectionrate}
are written in terms of the normalized expansion radius $R \equiv
r/r_L$.  We consider the acceleration to effectively begin at
normalized radius $R_0 = r_0/r_L$, where the outflow's Lorentz factor
is $\Gamma_0$, the comoving magnetic field is $\mathcal{B}_0^{'} $,
and the width of the current sheet is $\Delta_0 = 0$. The density at
$R_0$ is $n_2^{'} = n_0^{'}$ and the enthalpy is $h_0^{'} = m_p c^2
n_0^{'}$ (initially cold plasma).

The continuity Equation \ref{eq:continuityderivative} and energy-flux
Equation \ref{eq:energyderivative} take the form
\begin{align}
\beta \Gamma R^2 (n_1^{'} \Delta + (1-\Delta) n_2^{'}) & = \beta_0
\Gamma_0 R_0^2 n_0^{'}, \label{eq:continuityequationexpanded}\\
\beta \Gamma^2 R^2 \left [ h_1^{'} \Delta +(1-\Delta) \left ( h_2^{'}
  + \frac{\mathcal{B^{'}}^2}{4 \pi} \right ) \right ] & = \beta_0
\Gamma_0^2 R_0^2 \left ( \frac{\mathcal{B^{'}}_0^2}{4\pi} + m_p c^2
n_0^{'} \right ). 
\label{eq:energyequationexpanded}
\end{align}
The flux freezing condition (Equation \ref{eq:fluxfreezingderivative})
is simplified as
\begin{align}
\frac{\mathcal{B^{'}} }{ R n_2^{'} } = \frac{\mathcal{B}_0^{'} }{ R_0
  n_0^{'} }, \label{eq:fluxfreezingexpanded}
\end{align}
and it is used to eliminate $\mathcal{B}^{'}$ from the two previous
equations.

The complete derivation of the equations for $\Delta$ and $\sigma$ in
the first order approximation is somewhat technical, and presents only
a minor physical insight. Therefore, it is fully presented in
Appendix D, while in this section we provide the main results and
their physical interpretation.
In the zeroth order approximation in the small quantities
$\Delta$, $\sigma^{-1}$ and $\Gamma^{-2}$, the set of equations
becomes degenerate. Therefore, in order to obtain a solution, one needs to
solve the equations to first order in these quantities. Instead of working with equations containing both the zeroth and the first order terms, we manipulated them such that we eliminated the zeroth order terms, which yields a simple equation \ref{eq:D62}
\begin{align}
\Delta\left(2\frac{n'_1}{n'_2}-\frac { 3}{2}-\frac{(\epsilon_1^{'}-\epsilon_2^{'})\Gamma}{\Gamma_{\rm max}m_pc^2n'_2}\right)=
\frac{h_2^{'}\Gamma}{\Gamma_{\rm max}m_pc^2n'_2}, \label{eq:energyfluxfinal}
\end{align}
where $\Gamma_{\rm max} = \Gamma_0 \sigma_0$ is the maximum Lorentz
factor that can be achieved, if all the magnetic energy is converted
to bulk kinetic energy. In this equation, the right-hand side is small as $\Gamma/\Gamma_{\rm max}$ whereas the left-hand side is small as $\Delta$.

This equation may be supplemented by the zeroth order continuity equation
\begin{align}
R^2\Gamma n'_2=R_0^2\Gamma_0n'_0.
\end{align}
In the derivation of Equation
\ref{eq:energyfluxfinal}, it is explicitly assumed that the outflow
reached radius $R$ that is much larger than the fast magnetosonic
radius $R_{\rm fms}$. The fast magnetosonic radius corresponds to the
radius at which the outflow speed equals the speed of the fast magnetosonic wave, which
is given by $\beta_{\rm fms}^2 = (h^{'} s^2 + {\mathcal{B}^{'}}^2/(4\pi))/(h^{'}+{\mathcal{B}^{'}}^2/(4\pi))$,
where $s=\sqrt{dp /d\epsilon}$ is the sound speed. 
At radii $R \gg R_{\rm fms}$, one finds $\Gamma \gg \Gamma_{\rm max}^{1/3}$.
This can be easily understood in the highly magnetized regime.
Using the flux freezing condition and the zeroth order continuity equation, the Lorentz factor of the fast magnetosonic wave
is given by $\Gamma_{\rm fms}^2  \equiv 1/(1-\beta_{\rm fms}^2) =  (1 + {\mathcal{B}^{'}}^2/(4\pi h^{'}))/(1-s^2) \sim (9/8) {\mathcal{B}^{'}}^2/(4\pi n^{'}_2 m_p c^2) \sim \sigma_0 \Gamma_0/\Gamma $, where we explicitly assumed $ s^2 = 1/3$.
Since $\Gamma$ increases with radius while $\Gamma_{\rm fms}$ decreases with radius, at radii $R \gg R_{\rm fms}$ one finds $\Gamma \gg \Gamma_{\rm fms}$, and the condition $\Gamma^2 \gg \Gamma_{\rm fms}^2$ can be written as $\Gamma^3 \gg \Gamma_{\rm max}$.

The reconnection rate (Equation \ref{eq:reconnectionrate}) and the
entropy equation admit the forms
\begin{align}
\pder[\Delta \left( \frac{n_1^{'}}{n_2^{'}} -1\right )]{R} = \frac{A}{\Gamma^2}, \label{eq:reconnectionratefinal}
\end{align}
and 
\begin{align}
\frac{p_1 \Delta + (1-\Delta)p_2}{m_p c^2 n_2^{'} R^2 } \pder[\beta
  \Gamma R^2]{R} + \frac{1}{m_p c^2 n_2^{'} R^2 } \pder[\beta \Gamma
  R^2 (\epsilon_1^{'} \Delta + (1-\Delta) \epsilon_2^{'}]{R} \nonumber
\\ = c \Gamma_{\rm max} \pder [\Delta \left ( \frac{n_1^{'}}{n_2^{'}}
  -\frac{1}{2} \right )]{R},
~~~~~~~~~~~~~~ \label{eq:heatbalacecondition}
\end{align}
(see Equations \ref{eq:appendix_entropyexpanded} --
\ref{eq:appendix_heatbalacecondition} in Appendix D).

In deriving Equations \ref{eq:energyfluxfinal} --
\ref{eq:heatbalacecondition}, there was no need to explicitly
introduce the radiation field. Radiation affects the dynamics by
contributing to the entropy $h^{'}$, energy density $\epsilon^{'}$ and
pressure $p$. Therefore, in order to solve these equations, one must
assume \textit{a priori} how radiation is distributed in the
outflow. We studied two limiting cases. First,
we assume that radiation can not diffuse through the current sheet,
and second, we consider the scenario in which the energy and entropy
carried by the photons are fully redistributed in the two regions. We present the full solution to the equations in Appendix E. Here, we briefly present the key results. In
section 4 below, we discuss the validity of these two limiting cases
for parameter regions characterizing GRB outflows.

\subsection{Case I: the heat remains in the reconnection layer}

If photons can not diffuse efficiently through the reconnection layer,
the heat produced by the reconnection remains in the current
sheet. The magnetized region remains cold, $p_2 \ll
{\mathcal{B}^{'}}^2/(4\pi)$ and $\epsilon^{'}_2 = m_p c^2 n_2^{'}$. In
this case the current sheet is hot, $\epsilon_1^{'} = 3p_1 + m_p c^2
n_1^{'} \sim 3p_1$, where $p_1$ is dominated by the radiation (the adiabatic index is $\hat \gamma = 4/3$).
 
The conservation equations admit the form (see Equations
\ref{eq:app_E1} -- \ref{eq:app_E3} in Appendix E)
\begin{align}
\frac{\Gamma}{\Gamma_{\rm max}} & = \Delta \left ( 2
\frac{n_1^{'}}{n_2^{'}} -3\right
), \label{eq:caseIenergy_finalequation} \\
\pder[\frac{\Gamma}{ \Gamma_{\rm max}} + \Delta ]{R} & =
\frac{2A}{\Gamma^2}, \\
2 \pder[\Delta]{R} - \frac{\Delta}{\Gamma} \pder[\Gamma]{R} +
\frac{\Delta}{R} & = \pder[\frac{n_1^{'} \Delta}{n_2^{'}}]{R}.
\end{align}
These equations are satisfied by the ansatz $\gamma \propto \Delta
\propto R^{1/3}$ and $n_1^{'}/n_2^{'} = const$. Substituting and
looking for the coefficients, the solution is
\begin{align}
\Gamma & = \left ( 5A \Gamma_{\rm max} R \right
)^{\frac{1}{3}} \label{eq:lorentz_factr_case1}, \\
\Delta & = \left ( \frac{AR}{25 \Gamma_{\rm max}^2}\right
)^{\frac{1}{3}} \label{eq:delta_case1}, \\
\frac{n_1^{'}}{n_2^{'}} & = 4. \label{eq:ratiodensitiescase1}
\end{align}

In order to obtain analytical expressions for the radial evolution of
the rest of the physical quantities, it is convenient to define a
radius $R_{0,{\rm I}}$ as the radius at which the asymptotic solution
of the Lorentz factor (given in Equation \ref{eq:lorentz_factr_case1})
is equal to the initial Lorentz factor, $\Gamma_0$. We interpret this
radius as the radius at which efficient acceleration by magnetic
reconnection begins (the subscript I represents case I - the heat remains
in the reconnection layer). Similarly, we define $R_{c,{\rm I}}$ as
the radius at which the entire magnetic energy is exhausted by
reconnection, and interpret this radius as marking the end of the
acceleration phase, which is the beginning of the coasting phase. From
Equation \ref{eq:lorentz_factr_case1} it follows immediately that
$R_{0,{\rm I}} = \Gamma_0^3/(5A \Gamma_{\rm max})$ and $R_{\rm c,{\rm
    I}} = \Gamma_{\rm max}^2 /(5A)$.

Using the first order continuity Equation \ref{eq:appendix_continuity1order}
combined with Equation \ref{eq:ratiodensitiescase1}, the radial
evolution of the density in the magnetized (cold) region in the range
$R_{0,{\rm I}} ... R_{\rm c,{\rm I}}$ is given by
\begin{align}
n_2^{'} = \frac{\Gamma_0 R_{0,{\rm I}}^2 n_0^{'}}{(5A \Gamma_{\rm
    max})^{\frac{1}{3}} R^{\frac{7}{3}} \left [ 1+3\left( \frac{AR}{25
      \Gamma_{\rm max}^2} \right )^{\frac{1}{3}}\right ]} \sim
\frac{\Gamma_0 R_{0,{\rm I}}^2 n_0^{'}}{(5A \Gamma_{\rm
    max})^{\frac{1}{3}} R^{\frac{7}{3}} },
\end{align}

The radial evolution of the comoving magnetic field is obtained using the 
flux freezing condition (Equation \ref{eq:fluxfreezingexpanded}), 
\begin{align}
\mathcal{B}^{'} = \frac{\mathcal{B}_0^{'}}{R_{0,{\rm I}} n_0^{'}}
\times R \times \frac{\Gamma_0 R_{0,{\rm I}}^2 n_0^{'}}{(5A
  \Gamma_{\rm max})^{\frac{1}{3}} R^{\frac{7}{3}} \left [ 1+3\left(
    \frac{AR}{25 \Gamma_{\rm max}^2} \right )^{\frac{1}{3}}\right ]}
\sim \frac{\Gamma_0 R_{0,{\rm I}} \mathcal{B}_0^{'} }{(5A \Gamma_{\rm
    max})^{\frac{1}{3}} R^{\frac{4}{3}} }.
\end{align}
Finally, the comoving temperature of the current sheet\footnote{By
  assumption, the magnetized region is cold.}  is obtained by assuming
that the internal energy density is dominated by radiation. Then, the
pressure balance condition is written as
\begin{align}
k_{\rm B} T^{'}_1 n_1^{'} + \frac{a_{\rm th} {T^{'}_1}^4}{3} \sim
\frac{a_{\rm th}{T^{'}_1}^4}{3} = \frac{\mathcal{B^{'}}^2}{8\pi}.
\end{align}
Therefore, $T^{'}_1 = (3\mathcal{B^{'}}^2/(8\pi a_{\rm
  th}))^{1/4}$. Here, $a_{\rm th} = 4\sigma_{\rm SB}/c$ is the
radiation constant, and $\sigma_{\rm SB}$ is the Stefan-Boltzmann
constant.

\subsection{Case II: the heat is redistributed by radiation in the magnetized region}
\label{sec:3.2}

In the limiting case II, the heat is assumed to be redistributed in
the magnetized region by radiation, which fills the \textit{striped
  wind} with energy density $\epsilon_{\rm rad}^{'}$.  This case
represents a scenario in which the optical depth of the outflow is
large, while the width of the current sheet is small enough to enable
photons to diffuse into the magnetized region. In this scenario, we
can approximate the temperatures both inside and outside current
sheets to be equal, $T_1^{'} = T_2^{'} = T^{'}$, and non-relativistic
$k_{\rm B} T_i^{'} \ll m_e c^2$, where $m_e$ is the mass of an
electron.  Hence, the energy densities and pressure in both regions
are given by $\epsilon_{1,2}^{'} = \epsilon^{'}_{\rm rad} +n_{1,2}^{'}
k_{\rm B} T^{'}+ m_p c^2 n_{1,2}^{'}$, and $p_{1,2} = n_{1,2}^{'}
k_{\rm B} T^{'} + (1/3)\epsilon_{\rm rad}^{'}$.  In the magnetized
region, the magnetic pressure dominates the gas and radiation pressure
${\mathcal{B}^{'}}^2/(8\pi) \gg p_2$. We expect the density inside the
current sheet to be very large, $n_1^{'} \gg n_2^{'}$, in order to
balance the magnetic pressure $n_1^{'}k_{\rm B} T^{'} =
{\mathcal{B}^{'}}^2/(8\pi)$.  In addition, it can be checked \textit{a
  posteriori} that radiation dominates the thermal energy density. Therefore,
$\epsilon_{1,2}^{'} = \epsilon^{'}_{\rm rad} + m_p c^2 n_{1,2}^{'}$.

The conservation Equations (\ref{eq:energyfluxfinal} --
\ref{eq:heatbalacecondition}) are simplified to
\begin{align}
\left ( 1+ \frac{4 \epsilon_{\rm rad}^{'}}{3 m_p c^2 n_2^{'}} \right
)\frac{\Gamma }{\Gamma_{\rm max}} & = \frac{2 n_1^{'}
  \Delta}{n_2^{'}}, \\
\pder[\frac{n_1^{'} \Delta}{n_2^{'}}]{R} & = \frac{A}{\Gamma^2}, \\
\pder[\frac{\epsilon_{\rm rad}^{'}}{m_p n_2^{'}}]{R} + \frac{1}{3}
\frac{\epsilon_{\rm rad}}{m_p n_2^{'}} \frac{1}{ \Gamma R}
\pder[\Gamma R^2]{R} & = \frac{\Gamma_{\rm max}}{\Gamma}
\pder[\frac{n_1^{'} \Delta}{n_2^{'}}]{R}.
\end{align}
These equations are satisfied by the ansatz $\Gamma \propto n_1^{'}
\Delta / n_2^{'} \propto R^{1/3}$, and $\epsilon_{\rm rad}^{'}/(m_p
c^2 n_2^{'})= const.$ Substituting and looking for the coefficients,
the solution is
\begin{align}
\Gamma & = \left ( \frac{30}{7}A \Gamma_{\rm max} R\right )^{\frac{1}{3}}, \label{eq:30} \\
\frac{\Delta n_1^{'}}{n_2^{'}} & = \left ( \frac{147 A R}{100
  \Gamma_{\rm max}^2}\right )^{\frac{1}{3}}, \label{eq:ratiodeltan1n2}
\\
\frac{\epsilon_{\rm rad}^{'}}{m_p c^2 n_2^{'}} & =
\frac{3}{10}. \label{eq:temperaturecase2}
\end{align}

Comparing Equations \ref{eq:lorentz_factr_case1} and \ref{eq:30}, one
finds that difference in the asymptotic Lorentz factor between the two
limiting cases I and II is very small, and is only $(35/30)^{1/3} -1 \sim
5.3$\%, independent on the wind parameters. This result is unexpected,
and somewhat counter-intuitive. 

In analogy to the discussion that followed Equation
\ref{eq:ratiodensitiescase1}, we define the limiting radii $R_{0,{\rm
    II}}$ and $R_{\rm c,{\rm II}}$. From Equation \ref{eq:30}, we find
$R_{0,{\rm II}} = 7\Gamma_0^3 / (30 A \Gamma_{\rm max})$ and $R_{\rm
  c,{\rm II}} = 7 \Gamma_{\rm max}^2 /(30 A)$. In deriving an
analytical expression for the radial evolution of $n_2^{'}$ as well as
the other hydrodynamical quantities between $R_{0,{\rm II}}$ and
$R_{\rm c,{\rm II}}$, we use the zeroth order continuity equation
$\beta \Gamma R^2 n_2^{'} = \beta_0 \Gamma_0 R_0^2 n_0^{'}$ to
obtain 
\begin{align}
n_2^{'} = \frac{\Gamma_0}{\Gamma} \frac{R_{0,{\rm II}}^2}{R^2}
n_0^{'}. \label{eq:n2case2}
\end{align}
Using the flux freezing condition (Equation
\ref{eq:fluxfreezingexpanded}), the radial evolution of the comoving magnetic
field can be obtained:
\begin{align}
\mathcal{B}^{'} = \frac{\Gamma_0}{\Gamma} \frac{R_{0,{\rm II}}}{R} \mathcal{B}_0^{'}.
\end{align}
The comoving temperature is found by using Equation
\ref{eq:temperaturecase2} and the expression for $n_2^{'}$
\begin{align}
T^{'} = \left (\frac{\epsilon_{\rm R}^{'}}{a_{\rm th}} \right
)^{\frac{1}{4}} = \left ( \frac{3}{10a} m_p c^2 n_0^{'} \Gamma_0
R_{0,{\rm II}}^{2} \right )^{\frac{1}{4}} \left ( \frac{1}{\Gamma
  R^2}\right )^{\frac{1}{4}}.
\end{align}
The pressure balance condition allows to obtain $n_1^{'}$  
\begin{align}
n_1^{'} = \frac{\Gamma_0^2 R_{0,{\rm II}}^2 \mathcal{B^{'}}^2_0}{8\pi
  k_B \left [ \frac{3}{10a} m_p c^2 n_0^{'} \Gamma_0 R_{0,{\rm
        II}}^2\right ]^{\frac{1}{4}}} \frac{1}{\Gamma (\Gamma
  R^2)^{\frac{3}{4}}}, \label{eq:n1case2}
\end{align}
and the evolution of the current sheet width $\Delta$ is calculated
using Equations \ref{eq:ratiodeltan1n2}, \ref{eq:n2case2} and
\ref{eq:n1case2} to be
\begin{align}
\Delta = \frac{n_0^{'}}{\Gamma_0 \mathcal{B^{'}}_0^2} 8\pi k_B \left [
  \frac{3}{10a} m_p c^2 n_0^{'} \Gamma_0 R_{0,{\rm II}}^2\right
]^{\frac{1}{4}} \frac{\Gamma^{\frac{3}{4}}}{R^{\frac{1}{2}}} \left (
\frac{147 A R}{100 \Gamma_{\rm max}^2} \right
)^{\frac{1}{3}}. \label{eq:delta_case_II}
\end{align}
Using the scaling laws derived above for $\Gamma$, we find that the
current sheet width grows as $\Delta \propto R^{1/12}$, much slower
than in case I. This further implies that the density ratio
$n_1'/n_2'$ increases with radius, as $n_1'/n_2' \propto
R^{1/4}$. At least qualitatively, this increase in the density ratio
with radius can be understood as following from the requirement of pressure
balance in both sides of the current sheet, and the fact that energy
and entropy constantly ``leaks'' outside of the reconnection layer by
photon diffusion.

\subsection{Numerical integration}
\label{sec:3.3}

\begin{figure}
\centering
\begin{tabular}{cc}
\includegraphics[width=0.49\textwidth]{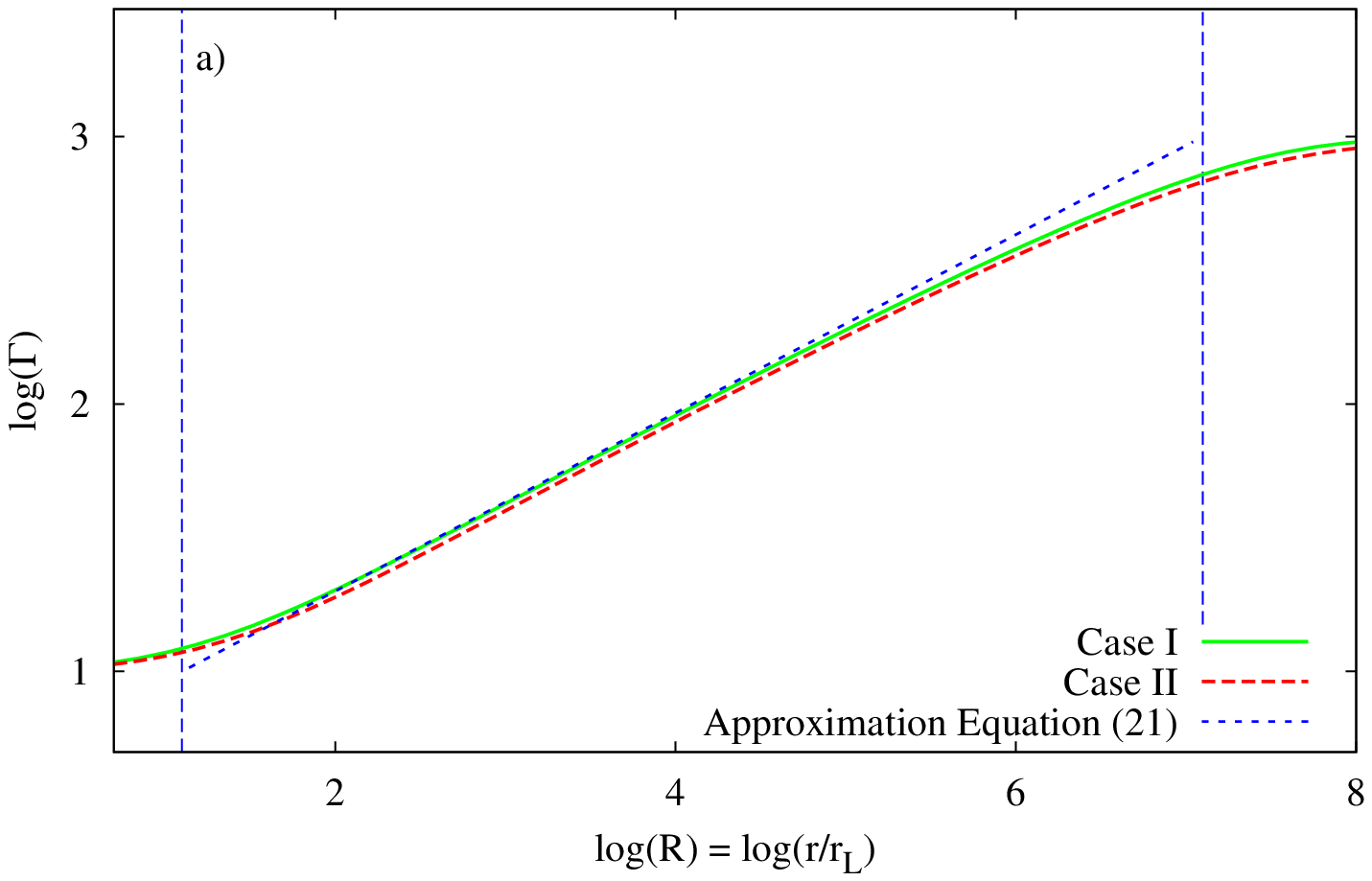} & \includegraphics[width=0.49\textwidth]{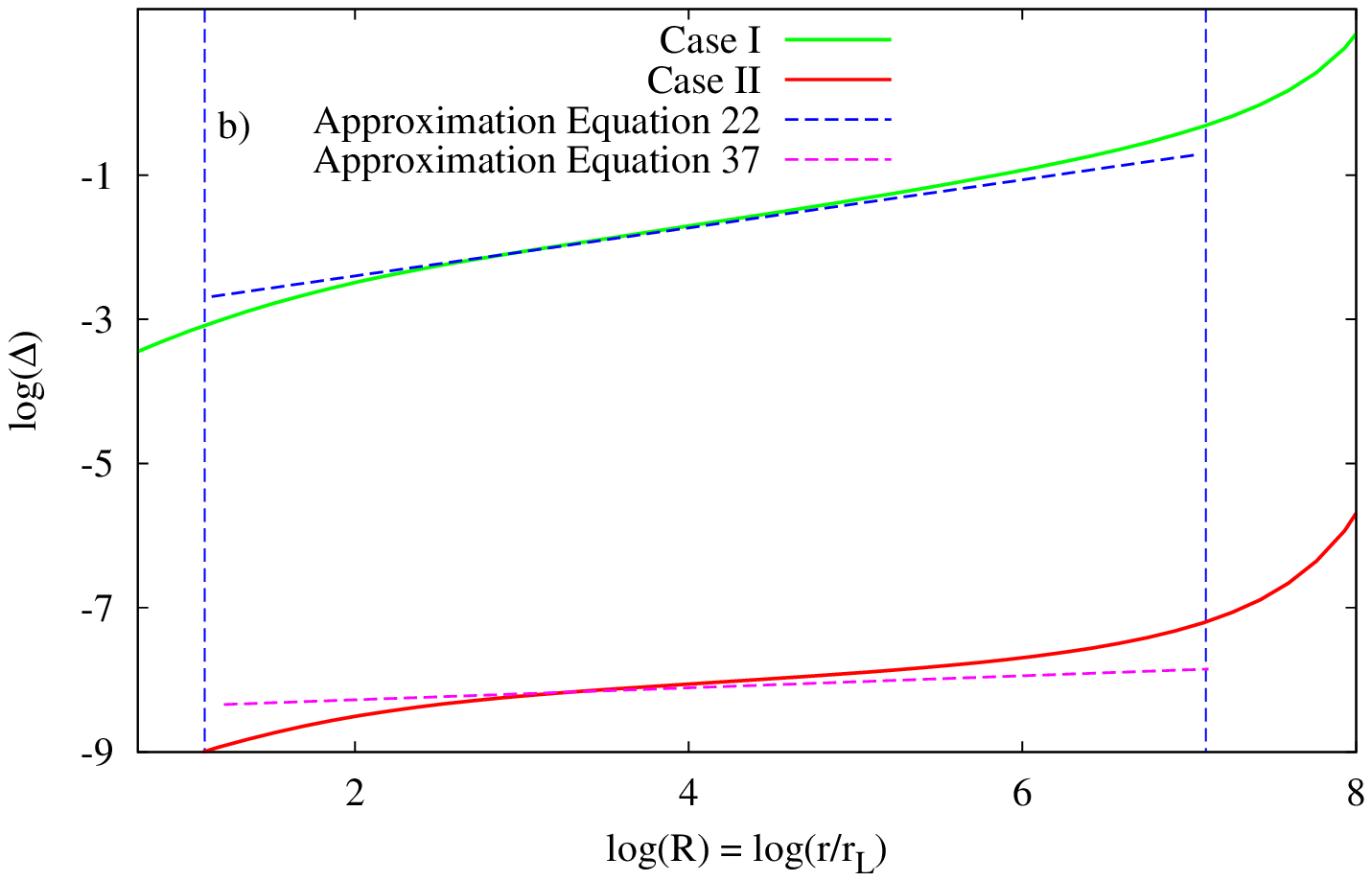} \\ 
\includegraphics[width=0.49\textwidth]{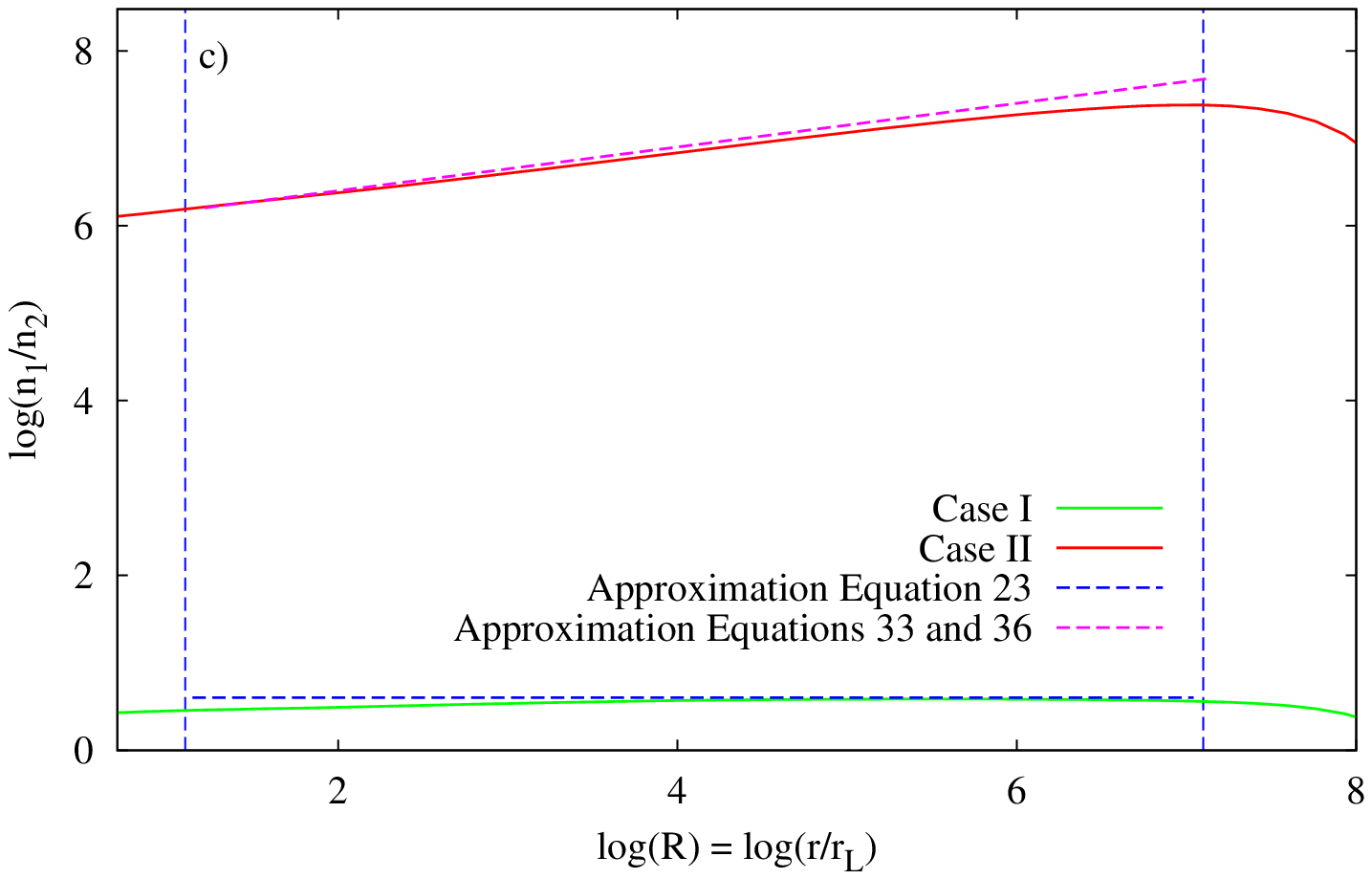} & \includegraphics[width=0.49\textwidth]{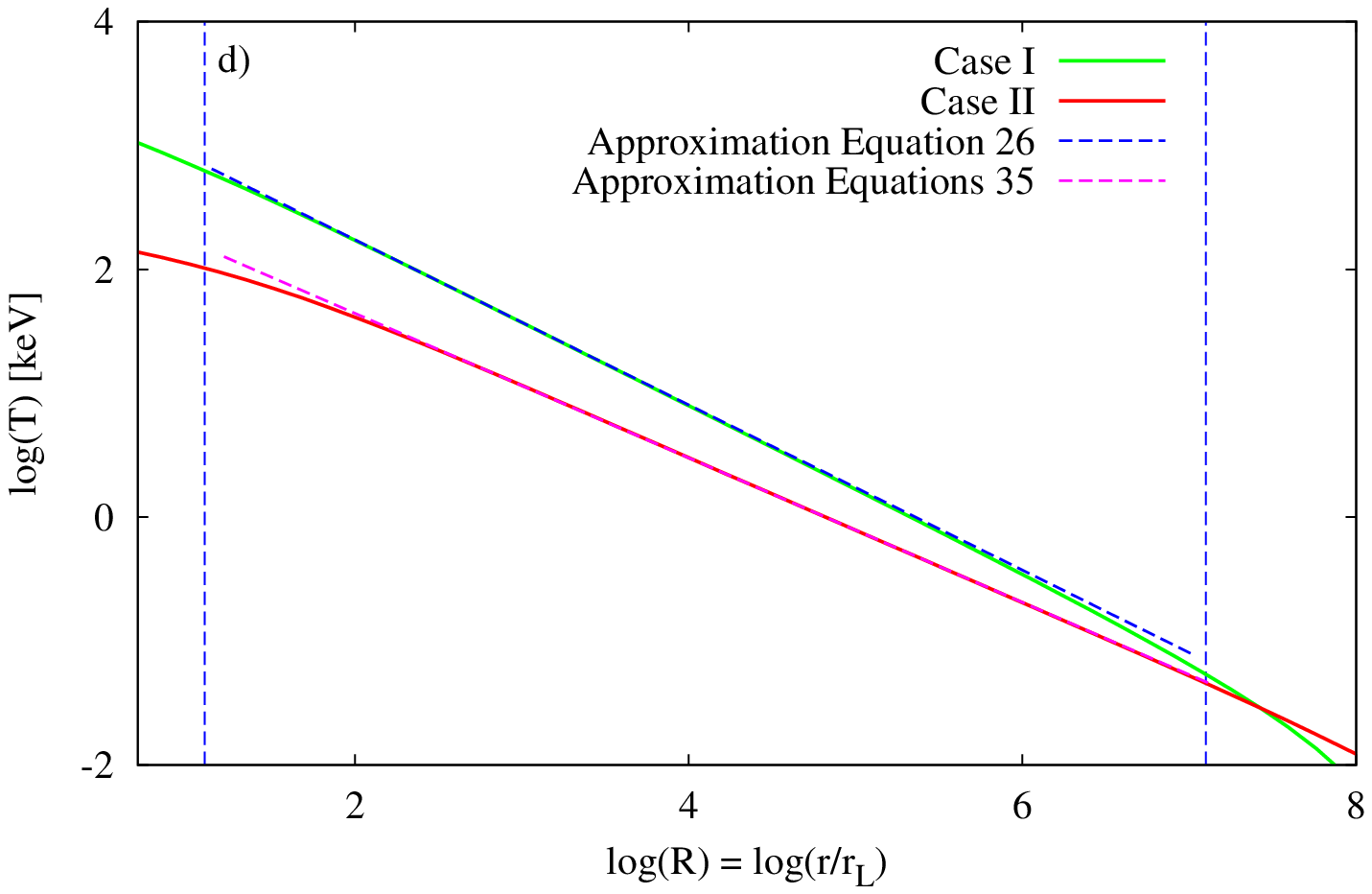}
\end{tabular}
\caption{From top left to bottom right, radial evolution of the Lorentz factor $\Gamma$, of the current sheet width $\Delta$, of the ratio of densities $n_1^{'}/n_2^{'}$, and of the temperature $T^{'}$. The blue and purple dashed lines represent the approximations derived in Section \ref{sec:approxsol}. Finally, the vertical blue dashed lines represent $R_{0,\rm I}$ and $R_{\rm c,\rm I}$. For clarity, we do not display the corresponding value for case II as they are respectively close to $R_{0,\rm I}$ and $R_{\rm c,\rm I}$.}
\label{fig:1}
\end{figure}

In this section, we obtain full numerical solution to the conservation
equations.
We present in Figure
\ref{fig:1} the radial evolution of $\Gamma$, $n_1^{'}/n_2^{'}$,
$\Delta$ and $T^{'}$ for cases I and II, together with the asymptotic
solutions derived above. In solving the equations, we chose fiducial
parameters $r_0 = r_L =10^6$ cm\footnote{Note that in the asymptotic regime, the solution is independent of $r_0$.}, $\sigma_0 = 100$, $\Gamma_0 = 10$ and $\epsilon  = 0.1$, corresponding to
$(\epsilon \Omega)_3 = 3$. With these parameters, the maximum Lorentz factor is $\Gamma_{\rm max} = \Gamma_0 \sigma_0 = 1000$.

Figure \ref{fig:1} clearly shows that the flow is very well described by our asymptotic solution in the intermediate zone and that the solution is not sensitive to
uncertainties in the initial conditions, in particular to the exact
initial values of $\Delta_0$, $T_0$ and $r_0$. In other words, the
flow at $r\gg r_0$ is fully described by $\Gamma_{\rm max} =\sigma_0\Gamma_0$ and
$(\epsilon \Omega)$ only, provided that the thermal energy is
initially much smaller than the magnetic energy as well as the rest
mass energy density.

The top panel a) of Figure \ref{fig:1} displays the radial evolution
of the Lorentz factor for cases I and II. For clarity, only the
asymptote given by Equation \ref{eq:lorentz_factr_case1} is
presented. After an initial coasting period, the outflow accelerates
with the Lorentz factor increasing proportionally to $R^{1/3}$, in
very good agreement with the asymptotic solutions. At larger radii,
after all the magnetic energy is exhausted, the outflow coasts. It is
clearly seen that the Lorentz factor is not strongly influenced by the
assumption on the evolution of the radiation field.

The width of the current sheet $\Delta = l_{\rm cs}/l_0$ is shown in
the second panel b) of Figure \ref{fig:1}. We find a good agreement
with the asymptote for case I, given by Equation
\ref{eq:delta_case1}. In case II, small deviations between the
numerical solution and the asymptote given by Equation
\ref{eq:delta_case_II} is seen, which can be explained by the use
of zeroth order approximation of the continuity equation. As expected,
when the current sheet is not supported by the radiation pressure, its
width is smaller by several order of magnitude while its density is
larger by several order of magnitude relative to case I.

The third panel c) of Figure \ref{fig:1} represents the ratio of the
density in the current sheet to that inside the magnetized region,
$n_1'/n_2'$. The agreement with the asymptotes derived in the previous
section is very good. In case I, the magnetic pressure is supported by
the thermal energy in the current sheet. As a result, its density does
not need to be much larger than that in the magnetized region, and the
ratio stays constant. On the other hand, when heat and entropy of the
current sheet can be redistributed by photon diffusion and scattering,
the magnetic pressure is supported by the thermal pressure of the
particles, implying a very large density in the current sheet.

Finally, the radial evolution of the current sheet comoving
temperature (which is equal to that of the whole outflow in case II)
is displayed in the fourth panel d) of Figure \ref{fig:1}. In both
cases the temperature decreases; and, with the exception of the initial stage, remains non-relativistic. Once again the agreements with the
asymptotes derived in the previous section are very good.

\begin{figure}
\centering
\includegraphics[width=0.50\textwidth]{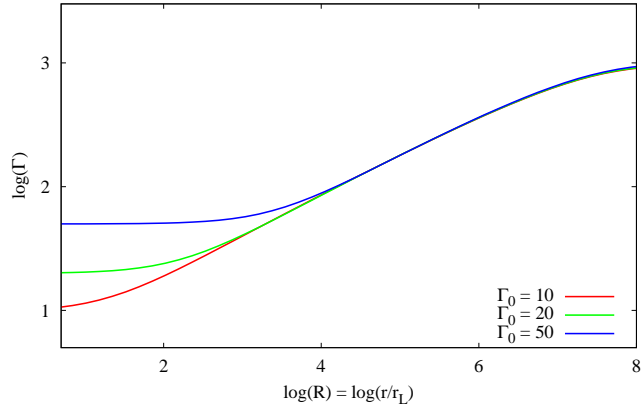} 
\caption{Radial evolution of the Lorentz factor for initial conditions $\Gamma_0= 10, 20, 50$ and same $\Gamma_{\rm max}$ in case II. After a transition corresponding to adjustments from the initial conditions, all evolutions follow the asymptote given by Equation \ref{eq:30}. }
\label{fig:2}
\end{figure}

To further illustrate that the evolution is solely sensitive to
the values of $\Gamma_{\rm max}$ and $(\epsilon \Omega)$, Figure \ref{fig:2} shows the radial
evolution of the Lorentz factor in case II for several values $\Gamma_0$,
but same $\Gamma_{\rm max} = \sigma_0 \Gamma_0$, showing that the solutions
approach the same asymptotic evolution provided that $\Gamma_{\rm max}$ is the same. Similar results hold for 
the other quantities.

In the next section, we study \textit{a posteriori} the validity of
the assumptions made in deriving cases I and II equations, and their
applicability to the GRB environment.

\section{Transitions radii in the striped wind scenario}
\label{sec:4}

\subsection{The photospheric radius}
\label{sec:4.1}

The derivations presented above are valid as long as the radiation and
matter are strongly coupled, namely below the photosphere. In
calculating the photospheric radius, one can use the general
expression of the optical depth for a photon emitted at radius $r$ on
the line of sight first derived by \citet{ANP91},
\begin{align}
\tau (r) = \sigma_{\rm T} \int_R^{\infty} \Gamma n^{'} (1-\beta) ds. \label{eq:opacity}
\end{align}
Here, $\sigma_{\rm T}$ is the Thompson cross-section, and the
integration is performed over the photon path. In deriving the
analytical expression below, we assume that the photospheric radius is
below the coasting radius, namely $r_{\rm ph}/r_L < R_{\rm c,I,II}$. As we show
below, this is the case for parameters characterizing GRB outflow.
The contribution of the current sheet to the total opacity can be neglected since only a small fraction of the matter is contained in the current sheet during the accelerating phase.
Using first order expansion $(1-\beta) \simeq 1/(2\Gamma^2)$ one finds 
\begin{align}
\tau(r) = \sigma_T \frac{\Gamma_0 R_{0,i}^2 n_{0,i}^{'}}{2
  \mathcal{A}_i} \frac{3}{5r^{\frac{5}{3}}},
\end{align}
where the index $i = {\rm I, II}$ for the two cases considered
above. Furthermore, in case I, $\mathcal{A}_i \equiv \mathcal{A}_{\rm
  I} =( 5 A \Gamma_{\rm max} )^{\frac{2}{3}} $, while in case II,
$\mathcal{A}_i \equiv \mathcal{A}_{\rm II} = \left ( 30 A \Gamma_{\rm
  max} /7 \right )^{\frac{2}{3}}$. The photospheric radius is, by
definition, the radius at which $\tau(r_{\rm ph})=1$, from which one finds
\begin{align}
r_{\rm ph} & = \left [\frac{3 \sigma_{\rm T} \Gamma_0 R_{0,i}^2 n_{0,i}^{'}}{10 \mathcal{A}_i} \right ]^{\frac{3}{5}}  = \left\{ \begin{array}{l l}
 6.8 \times 10^{11} ~ L_{52} ^{\frac{3}{5}} (\epsilon \Omega)_3^{-\frac{2}{5}} \Gamma_{\rm max,3}^{-1} {\rm ~ cm} & {\rm case~ I,} \\
7.0 \times 10^{11} ~ L_{52} ^{\frac{3}{5}} (\epsilon \Omega)_3^{-\frac{2}{5}} \Gamma_{\rm max,3}^{-1} {\rm ~ cm}& {\rm case~ II}.
\end{array} \right.
\label{eq:r_ph}
\end{align}
Here and below, $X = 10^n X_n$ in cgs units. We thus conclude that the
difference between the photospheric radii is small, as it is mainly
due to the difference in the Lorentz factor. Since the parametric
dependence is the same in both cases, we drop the distinction between
case I and case II when referring to the photospheric radius from here
on. In addition, the photospheric radius is almost two orders of magnitude 
smaller than the coasting radius for the fiducial parameters of our model.

\subsection{Transition between the regimes I and II}
\label{sec:4.2}

We considered the flow in two cases: (I) the radiation is locked in the current sheet and (II) the radiation escapes freely the current sheet filling the whole space. Applying our results to GRBs, we now check when our assumptions are valid.

In case I, the key assumption is that photons remain in the current sheet. It requires that the rate of diffusion through the current sheet be small. This condition ceases to be fulfilled at the radius $r_{\rm D, I}^{\Delta}$ above which radius, diffusion through the current sheet is efficient.  
When Compton scattering dominates, the diffusion time through the current sheet is given by\footnote{The factor of two at the denominator comes from the fact that energy diffuses from the center of the current sheet towards both boundaries.}
\begin{align}
t_{\rm D}^{\Delta} = \frac{\sigma_T n_1^{'} (\Gamma \Delta l_0)^2}{2 c}, \label{eq:appendixGdiffusiontimeDelta}
\end{align}
while the diffusion time through the magnetized region is obtained by replacing $\Delta$ by $1-\Delta$. 
Equating the diffusion time to the expansion time $t_{\rm exp}=r/\Gamma c$ gives for the current sheet
\begin{align}
r_{\rm D, I}^{\Delta} & = 6.05 \times 10^{10} ~ L_{52}^{\frac{3}{5}}
(\epsilon \Omega)_3^{\frac{4}{5}} l_{0,7}^{\frac{6}{5}}
\Gamma_{\rm max,3}^{-1} {\rm ~
  cm}.\label{eq:diffusion_case1_delta} 
\end{align}
Above $r_{\rm D, I}^{\Delta}$, energy leaves the current sheet, taking away heat and entropy, which are (partially) redistributed in the magnetized region by Compton scattering, since $r_{\rm D, I}^{\Delta} < r_{\rm ph}$.

Case II is defined by two conditions. First, it is assumed that the energy deposited by reconnection in the current sheet can be redistributed in the magnetized region. It implies that diffusion through the current sheet is efficient. Second, the temperature in the current sheet and in the magnetized region are assumed to be similar $T_1^{'} \sim T_2^{'}$, such that the pressure balance condition is satisfied by the gas pressure rather than the radiation pressure. We point out that in practice because energy is continuously deposited by reconnection in the current sheet, the average temperature of the current sheet $\bar{T}_1^{'}$ is always larger than the average temperature in the magnetized region $\bar{T}_2^{'}$, as long as the opacity of the current sheet be larger than the unity. Still for $T_1^{'} \gtrapprox T_2^{'}$ at the boundary of the current sheet, $\Delta p_{\rm g} \gg \Delta p_{\rm rad}$, and the conditions for case II are satisfied.

In the regime II, the plasma density in the current sheet is so high that the free-free absorption/emission becomes important. 
Using the Rosseland mean for free-free emission $\alpha_{\rm R}^{\rm ff}$, given by Equation 5.20 of \cite{RL79},
the ratio of free-free opacity to photon scattering opacity is given by
\begin{align}
\frac{\alpha_{\rm R}^{\rm ff}}{\sigma_T n_e^{'}} = 0.24 ~ \Gamma_{\rm max,3}^{\frac{5}{6}} (\epsilon \Omega)_3^{-\frac{7}{24}} r^{-\frac{1}{24}} L_{52}^{-\frac{1}{8}}.
\end{align}
One sees that the free-free opacity remains less than the scattering opacity, the photon diffusion time is determined by the Thomson scattering. However, free-free absorption/emission efficiently thermalized radiation provided the condition
\begin{align}
\sqrt{\alpha^{\rm ff}_{\rm R}\sigma_Tn'_e}\Delta \Gamma l_0 >1,
\end{align}
is fulfilled \citep{Sha72_2}. Substituting the parameters of the flow, one finds that the radiation from the current sheet is thermalized up to the distance
\begin{align}
r^{\Delta}_{\rm ff, II}= 2.2 \times 10^{11} L_{52}^{\frac{5}{9}}
(\epsilon \Omega)_3^{\frac{1}{9}} l_{0,7}^{\frac{16}{27}}
\Gamma_{\rm max,3}^{-\frac{20}{27}} {\rm ~
  cm}. \label{eq:caseII_thermlization}
\end{align}
In the context of GRB, this might have profound effects on the emitted spectrum at the photosphere. In particular, it has the potential of solving the soft photon problem (see section \ref{sec:52} below). A detailed analysis of photon production and Comptonization below the photosphere in the striped wind scenario is out of the scope of the current paper, but it will be presented in a future publication.

Above this radius, the condition for case II, namely efficient energy diffusion through the current sheet is satisfied. We mention that for equal average temperatures, $\bar{T}_1^{'} \sim \bar{T}_2^{'}$, one also needs to require efficient diffusion through the magnetized region. This condition is satisfied above the radius  
\begin{align}
r_{\rm D, II}^{1-\Delta} = 8.0 \times 10^{11} ~ L_{52}^{\frac{3}{7}} l_{0,7}^{\frac{6}{7}} (\epsilon \Omega)^{\frac{2}{7}} \Gamma_{\rm max,3}^{-\frac{1}{7}} {\rm ~cm}
\end{align} 
obtained by using Equation \ref{eq:appendixGdiffusiontimeDelta} with $\Delta \rightarrow 1-\Delta$. We point out that $r_{\rm D, II}^{1-\Delta}>r_{\rm ph}$, above which Compton scattering ceases. Therefore, the energy is never fully homogenized in the magnetized region.

In the above computations, we presented two limiting cases: case I in which the radiation is locked in the current sheet, and case II, in which radiation can stream efficiently through the current sheet.
At the formation of the reconnection layer, radiation is always trapped inside of it. Below the diffusion radius $r_{\rm D, I}^{\Delta}$, given by Equation \ref{eq:diffusion_case1_delta}, radiation remains in the current sheet. Therefore, the dynamics follows the limiting case I until this radius.

Only at radii $r > r_{\rm D, I}^{\Delta}$, energy can diffuse through the reconnection layer, and the contribution of radiation to the pressure balance condition decreases. As a result, the current sheet shrinks and its density increases. 
At $r_{\rm D, I}^{\Delta}$, the opacity of the current sheet is larger than the unity for parameters characterizing limiting case II. It implies that the average temperature of the current sheet $\bar{T}_1^{'}$ is larger than the temperature of the magnetized region $\bar{T}_2^{'}$ at these radii. Yet, because the temperature is weakly dependent on the optical depth\footnote{The relation between temperature and optical depth is easily obtained by writing the homogeneous diffusion equation without source term as $(\partial^2 T^4)/(\partial \tau^2) = 0$, from which $T^{'} \propto \tau^{1/4}$.}, with $T^{'} \propto \tau^{1/4}$, one could expect that the dynamics will not be significantly modified. Moreover, as it was shown above, the radiation energy density within the magnetized region remains inhomogeneous, contrary to the assumptions of the case II. This means that our case II may be considered as an idealized limiting case. But taking into account that the dynamics of the flow is nearly the same on the limiting cases I and II, one can be sure that it remains the same even in the intermediate case.


\section{Discussion}

\subsection{Dynamics}

We find that the acceleration of the outflow is very similar in limiting cases I and II. The radial evolution of the Lorentz factor is mostly due to our choice of reconnection rate, given by Equation \ref{eq:reconnectionrate}. It constitutes the main assumption of our work.
As already pointed out, when studying pulsar wind nebulae, \cite{Cor90} and \cite{LK01} parametrize the width of the reconnection layer $\Delta$ by its proportionality with the Larmor radius, such that $\Delta \Gamma l_0 = \kappa k_{\rm B} T^{'}_1 / (e \mathcal{B}^{'})$, where $\kappa$ is a constant in the order of the unity\footnote{Both \cite{Cor90} and \cite{LK01} set $\kappa=2$.}. The main difference with the GRB scenario is the important baryon load expected from the progenitor, see \textit{e.g.} \cite{LE03}. In fact, reconnection happens when the charge density in the current sheet cannot sustain the required current\footnote{Or in other words, when the speed of the charge carriers becomes $c$.} \citep{Uso75}, which in the context of GRBs happens only at very large radius. 

The micro-physical details of reconnection happen on the length-scale of few tens of plasma skin
depth \citep{SS14}. Comparing the comoving sheet width and the plasma skin depth $\Lambda$
of the current sheet at the photosphere, we find for case II:
\begin{align}
l_{\rm cs}  = \Delta l_0 \Gamma & = 66 ~ L_{52}^{\frac{1}{2}} l_{0,7} (\epsilon \Omega)_3^{\frac{3}{4}}  \Gamma_{\rm max,3}^{-\frac{7}{4}}  {~ ~ \rm cm,}\\
\Lambda  =   \frac{c}{\sqrt{\frac{4\pi n^{'}_1 e^2}{m_p}}} & = 2.2 \times 10^{-4} ~ L_{52}^{\frac{1}{4}} \Gamma_{\rm max,3}^{-\frac{7}{8}} (\epsilon \Omega)_3^{-\frac{1}{8}} {~ ~ \rm cm.}
\end{align}
It is readily checked that $l_{\rm cs} \gg \Lambda$ at every radii. A similar result holds for case I as well.
This result therefore implies that the evolution of the striped wind is described by a macro-physical rather than micro-physical process.

\cite{SDD01} postulated the reconnection rate as given by Equation \ref{eq:reconnectionrate}, assuming that the alternating magnetic field annihilates with a rate close to $\sim 0.1$ of the Alfv\'en speed, which is comparable to the speed of light in highly magnetized outflows. Later, \cite{Lyu10} justified this assumption by showing that the reconnection is triggered by the self-sustained Kruskal-Schwarzschild instability. This instability makes the plasma ``drips out" of  current sheets as a result of its own acceleration, favouring the field reconnection and maintaining the plasma acceleration, hence the self-sustained characteristic of the instability. In the context of GRBs, reconnection is therefore the result of a macro-physical MHD instability.

The MHD instabilities will eventually destroy the \textit{striped wind.} This process was studied by \cite{Zra15} in the context of pulsar winds under the guise of force-free electrodynamics. In this work, it was found that the plasma instabilities become dominant once the causal contact is established between the stripes, which happens after a comoving fast magnetosonic time. The characteristic radius at which this happens, $r_{\rm tur}$,  can be calculated by comparing the comoving expansion time $r/(\Gamma c)$ to the crossing time of a stripe by a fast magnetosonic wave, $\Gamma l_0 / (\beta_{\rm fms} c)$, where $\beta_{\rm fms}$  is the speed of the fast magnetosonic wave, associated with the Lorentz factor $\Gamma_{\rm fms} = (\sigma_0 \Gamma_0 / \Gamma)^{1/2}$. Assuming $\beta_{\rm fms} \sim 1$ and case I, 
one obtains ${r_{\rm tur} = 7.1 \times 10^{11} l_{0,7}^3 (\epsilon \Omega)_3^2 \Gamma_{\rm max,3}^2}$cm, 
close to the photospheric radius. Note the very strong dependence on the parameters, and especially
on $l_0$. The turbulence is fully established and developed after a time four times longer than it takes the fast magnetosonic wave to cross one stripe \citep{Zra15}. It corresponds to the radius
$r=4.5 \times 10^{13} l_{0,7}^3 (\epsilon \Omega)_3^2 \Gamma_{\rm max,3}^2$cm.

Finally, we note that if all the magnetic energy were to be transformed to kinetic energy, the maximum Lorentz factor achieved by the outflow would be $\Gamma_{\rm max} = \sigma_0 \Gamma_0$. However, under the assumption of high radiative efficiency, \cite{DS02} explained that a large fraction of the total energy is carried away by radiation, substantially reducing the final Lorentz factor. It was recently shown by \cite{Pee16} that up to half of the total energy can be emitted, reducing the final Lorentz factor down to $\Gamma_{\rm max}/2$. In this paper, we discarded this effect as we were interested in the solution below the photosphere, where coupling between the radiation and particles is efficient, as required by our assumptions. A detailed description of the spectrum emitting at the photosphere and at larger radii will be given in a future work.

\subsection{Implications for the prompt emission of GRBs}
\label{sec:52}

The observed MeV peak energy in the spectrum of the prompt GRB emission can be attributed to thermalization processes inside the photosphere of the outflow \citep{EL00, MR00, RM05, TMR07, Bel13, VLP13}. This requires models of sub-photospheric dissipation to be able to produce photons efficiently. Indeed, in a scenario in which photons cannot be produced and if dissipation occurs, the photons already present in the outflow would share the dissipated energy. As their average comoving energy is  strongly increased, the observed spectral peak energy is shifted to large values. Assuming a luminosity-peak energy correlation, also referred to as Yonetoku correlation \citep{YMN04}, \cite{Bel13} and \cite{VLP13} found that the Lorentz factor of GRBs can not be larger than $\sim 10$ at the dissipation radius. This analysis is based on photon production processes, necessary to achieve thermalization and regulate the energy of the thermal peak. This result holds independently of both the content (baryonic, magnetic and thermal) and of the acceleration mechanism of the plasma, still unknown and highly debated \citep{ZP09,BGP15,BP15}.

Referring to the \textit{striped wind} scenario, \cite{VLP13} also considered the Bremsstrahlung rate when the plasma is clumped, \textit{i.e.} when over-dense regions exist. Indeed, because Bremsstrahlung depends on the density square, compressed regions could play an important role in producing photons. Yet, they had to assume the fraction of energy carried by Poynting flux. They further estimated the compression ratio to be around $10^6$ by satisfying the pressure balance condition with the  gas pressure.
In this paper, we consistently obtained the compression ratio, based on first principle MHD, and find that it agrees with the estimate of \cite{VLP13} for case II.

In case I, the compression ratio is small, being only 4. Therefore, the Bremsstrahlung process is not expected to be much more efficient than if no matter clumping was present. Using the same procedure as in \cite{VLP13} and \cite{Bel13}, we find that Bremsstrahlung freezes out at radius 
\begin{align}
\bar{r}_{\rm ff, I}^{\Delta} & =  8.0 \times 10^8 ~ \left (\frac{\bar{A}}{15} \right )^{\frac{3}{5}} L_{52}^{\frac{27}{40}} \Gamma_{\rm max,3}^{-\frac{29}{20}} (\epsilon \Omega)_3^{-\frac{1}{4}} ~ \text{cm,} \label{eq:rffstopcase1}
\end{align}
where $\bar{A}$ depends weakly on the temperature and is found to be nearly constant \citep{BP15}. This result is compatible with that of \mbox{\cite{VLP13}} and \mbox{\cite{Bel13}}. We note that in this situation, the outflow is unable to sustain the thermalization of the plasma as photons cannot be created (unless the outflow remains relatively slow with $\Gamma \sim 10$ at radius $r\sim 10^{10}$cm). Therefore, the emission of such an outflow is characteristic of the photon starvation scenario discussed in \cite{BP15}.

In case II however, 
the compression ratio is very large $\sim 10^{6}$ (as required by the pressure balance condition). As a result, Bremsstrahlung photon production is efficient at least up to $r_{\rm ff,II}^{\Delta} $ given by \ref{eq:caseII_thermlization}, which implies that the radiation from the current sheet be thermalized almost up to the photosphere.

\section{Conclusion}

In this paper, we presented a detailed analysis of the effects of radiation on the \textit{striped wind} model.
In the context of GRBs, this is required as the wind is initially optically thick due to the large baryon load. Previous studies of the dynamics of the \textit{striped wind} model in the context of GRBs \citep{DS02,Dre02} discarded the internal structure of the wind, preventing to study how radiation effects the dynamics of the plasma and its internal structure.

We presented an analytic asymptotic solution to the relativistic magneto-hydrodynamics equations by employing the short-wavelength approximation initially developed in \cite{LK01}. The solution depends on whether or not radiation can diffuse through the current sheet. Yet, the difference in  the evolution of the Lorentz factor between the two limiting scenarios is small, and only the current sheet width and density are substantially modified.

We explained that case I is initially relevant for wind parameters compatible with GRB physics. Above $r_{\rm D,I}^{\Delta}$, photons in the current sheet diffuse into the magnetized region. As a result, the radiation pressure of the current sheet drops, implying a decrease of its width, while its density increases to compensate for the magnetic pressure. Once this happens, the dynamics transitions to the case II scenario. The rate of photon production by the free-free process increases, implying that the current sheet be thermalized up to $\sim 10^{11}$cm, slightly smaller than the photosphere. This effect could provide a solution to the soft photon problem in GRBs.

To conclude, our results pave the way for studies dedicated to the effects of radiation on the dynamical evolution of a \textit{striped wind}.

\section*{Acknowledgements}

DB is supported by a grant from Stiftelsen Olle Engkvist Byggm\"astare. AP acknowledges support  by the European Union Seventh Framework Programme
(FP7/2007-2013) under grant agreement ${\rm n}^\circ$ 618499. YL is supported by the Israeli Science Foundation under the grant no 719/2014.

\appendix

\section{Relativistic radiative MHD equations of the \textit{striped wind}}

The relativistic radiation hydrodynamics equations in spherical coordinates are given by \textit{e.g.} \cite{Par04}.
In this appendix, the expressions are directly specialised to spherical symmetry, namely $\partial/\partial \theta=\partial/\partial \phi=0$. In the context of the striped wind, the stress energy tensor can be divided into its matter and field components, $T = T_{\rm g} + T_{\rm M}$.  The stress energy tensor of an ideal gas is $T_g^{\mu \nu} = h_{\rm g} U^\mu U^\nu + p_{\rm g} g^{\mu \nu}$. Here, $U^{\mu} = (\Gamma, \Gamma v , 0 ,0)$ is the 4-velocity, $\Gamma$ and $v$ are the Lorentz factor of the flow and the speed, $p_{\rm g}$ is the gas pressure, $h_{\rm g} = \epsilon_{\rm g}+p_{\rm g}$ is the proper gas enthalpy, and its proper energy density is $\epsilon_{\rm g}$. Finally, $g^{\mu \nu}$ is the Minkowski metric tensor.

The electro-magnetic component of the plasma is described by the electro-magnetic stress-energy tensor $T_{\rm M}^{\alpha \beta}=(1/4\pi)(F^{\alpha \gamma} F^{\beta}_{~\gamma}-(1/4)g^{\alpha \beta} F_{\gamma \nu} F^{\gamma \nu} )$, where $F^{\alpha \beta}$ is the Maxwell field tensor. In the context of the \textit{striped wind}, far from the light-cylinder $r_{\rm L}$, the laboratory frame magnetic field is toroidal $\vec{\mathcal{B}}=\mathcal{B}(r,t) \vec{\phi}$. It implies that the laboratory frame electric field and current density are $\vec{\mathcal{E}}= \mathcal{E}(r,t) \vec{\theta} $ and $\vec{j}= j(r,t) \vec{\theta} $. The only non-null components of the Maxwell tensor are $F^{02} = r \mathcal{E}$, $F^{12} = -r \mathcal{B}$, $F^{20} = -r \mathcal{E}$ and $F^{21}=r \mathcal{B}$.

The radiation stress energy tensor is $R^{\mu \nu} = \iint I(\vec n ,\nu) n^\mu n^\nu d\nu d\Omega$, where $I$ is the specific intensity of photons of frequency $\nu$ moving in the direction $\vec n$. We further assume that radiation is isotropic in the comoving frame. It implies that all the only non-vanishing components of $R^{\mu\nu}$ are on the diagonal, when it is expressed in the comoving frame. In addition, 
$ {\rm diag} (R^{\mu \nu}) =  (\epsilon_{\rm rad}^{'}, P^{'}_{\rm rr}, P^{'}_{\theta\theta}, P^{'}_{\phi \phi})$ where $\epsilon_{\rm rad}^{'}$ is the comoving energy density in radiation, and $P^{'}_{\rm rr} = P^{'}_{\theta\theta} = P^{'}_{\phi \phi} = \epsilon_{\rm rad}^{'} /3$.

The number density conservation $(nU^{\alpha})_{;\alpha}$ becomes
\begin{align}
\frac{\partial}{\partial t} (\Gamma n^{'}) + \frac{1}{r^2} \frac{\partial}{\partial r}(r^2 n^{'} v \Gamma)=0. \label{eq:appendixcontinuityeq}
\end{align}
By noting $G^{\mu}$ the radiation 4-force density, further assuming that there is no external force, the Bianci identity for the radiation on the one hand and matter plus magnetic component on the other hand can be separated. It comes
\begin{align}
T^{\mu \nu}_{~~;\nu} & = G^{\mu} & \text{matter and electromagnetic field}, \label{eq:bianci_matter} \\
R^{\mu \nu}_{~~;\nu} & = -G^{\mu} & \text{radiation.}
\end{align}
The energy equation is the zeroth component of the divergence of the stress energy tensor
\begin{align}
\pder[T^{00}]{t} + \frac{1}{r^2} \pder[r^2 T^{01}]{R} = G^0, \label{eq:appendix_energy}
\end{align}
where $G^{0}$ is the zeroth component of the radiation four-force. With $T^{00}_{\rm M} = (\mathcal{E}^2+\mathcal{B}^2)/(8\pi)$ and $ T_{\rm M}^{01} =  c \mathcal{E}\mathcal{B}/(4\pi)$, the energy equation reduces to
\begin{align}
\pder[h_{\rm g}^{'} \Gamma^2 - p_{\rm g}]{t}+\frac{1}{r^2}\pder[r^2 v \Gamma^2 h_{\rm g}^{'}]{r} + \pder[\frac{\mathcal{E}^2+\mathcal{B}^2}{8\pi}]{t} + \frac{1}{r^2} \pder[r^2 \frac{\mathcal{EB}c}{4\pi}]{r} = G^{0}.\label{eq:energy_equation} 
\end{align}
The entropy equation is obtained by projecting Equation \ref{eq:bianci_matter} along $U_{\mu}$ and by using ${T^{\mu\nu}_{M~;\nu} = -F^{\mu}_{~l}J^{l}}$, where $J=(cqn,0,j,0)^{\rm T}$ is the four-current. The equation can be simplified to:
\begin{align}
-n^{'}\Gamma \pder[\frac{h^{'}_{\rm g}}{n^{'}}]{t} - n^{'} \Gamma v \pder[\frac{h^{'}_{\rm g}}{n^{'}}]{r} + \Gamma \pder[p_{\rm g}]{t} +\Gamma v \pder[p_{\rm g}]{r} = j\Gamma(\mathcal{E}-\frac{v}{c} \mathcal{B})-\Gamma G^{\rm 0}+ \Gamma v G^{\rm 1}, \label{eq:entropy_appendix}
\end{align}
where we used Equation \ref{eq:appendixcontinuityeq} to simplify the terms in $r^2$. In the absence of radiation ($G^0 = G^1 = 0$), when specializing to the equation of state $p_g=(\hat{\gamma}-1)(\epsilon_g-n^{'}mc^2)$, where $\hat \gamma$ is the adiabatic index of the gas\footnote{The adiabatic index $\hat \gamma$ is for the gas only. The radiation is treated separately with adiabatic index $4/3$.} and $c$ the speed of light, one recovers the entropy equation A4 of \cite{LK01}:
\begin{align}
\frac{1}{\hat{\gamma}-1} \left [ \frac{\text{d}}{\text{d} t} p_{\rm g} - \frac{\hat{\gamma} p_{\rm g}}{n^{'}} \frac{\text{d}}{\text{d}t} n^{'}\right ] & = j(\mathcal{E}-\frac{v}{c} \mathcal{B}),
\end{align}
where the convective derivative is
\begin{align}
\frac{\text{d}}{\text{d}t} = \frac{\partial}{\partial t} + v \frac{\partial}{\partial r}.
\end{align}

We now turn to express the conservation equation for the radiation. In the laboratory frame the radiation stress energy tensor becomes
\begin{align}
R^{\mu \nu} = \left ( \begin{tabular}{cccc}
 $\null \epsilon_{\rm rad} = \frac{4}{3}\Gamma^2 \epsilon_{\rm rad}^{'} - \frac{\epsilon_{\rm rad}^{'}}{3}$  & $F^{\rm r} = \frac{4}{3}\Gamma^2 v \epsilon_{\rm rad}^{'}$  & $0$ & $0$ \\
 $\null F^{\rm r}$ & $P _{\rm rr } = \left( \frac{4}{3} \Gamma^2 -1\right )\epsilon_{\rm rad}^{'}$ & $0$  & $0$ \\ 
 $\null 0$ & $0$ & $P_{\rm \theta\theta }= P^{'}_{\rm \theta \theta}$ & $0$ \\ 
 $\null 0$ & 0 & 0 & $P _{\rm \phi \phi }= P_{\rm \phi \phi}^{'}  $
 \end{tabular} \right ).
\end{align}

The radiation energy equation is obtained by taking the zeroth component of the divergence of the radiation stress energy tensor
\begin{align}
\pder[\epsilon_{\rm rad}]{t}+\frac{1}{r^2} \pder[r^2 F^{\rm r}]{r} = -G^0, \label{eq:radiation_energy}
\end{align}
while the radiation momentum equation is obtained from the first component of the divergence of the radiation stress energy tensor
\begin{align}
\pder[F^{\rm r}]{t} + \pder[P_{\rm rr}]{r} + \frac{8}{3}(\Gamma^2 -1) \frac{\epsilon_{\rm rad}^{'}}{r} = -G^{1}.
\end{align}
These two equations give the expression of the radiation four-force appearing in Equations \ref{eq:energy_equation} and \ref{eq:entropy_appendix}.

The system is completed by Maxwell's equations. Ampere's law reads
\begin{align}
\frac{1}{r} \pder[r\mathcal{B}]{r} + \frac{1}{c} \pder[\mathcal{E}]{t}+\frac{4\pi}{c} j =0 .
\end{align}
Faraday's law is
\begin{align}
\frac{1}{r} \pder[r\mathcal{E}]{r} + \frac{1}{c} \pder[\mathcal{B}]{t} =0,
\end{align}
and Ohm's law 
\begin{align}
j=\sigma_c \gamma \left (\mathcal{E}-\frac{v}{c} \mathcal{B} \right).
\end{align}
Here $\sigma_{\rm c}$ is the conductivity of the plasma.
The system is closed by the ideal gas law and an equation of state:
\begin{align}
p_{\rm g} & = k_{\rm B} n^{'} T^{'}, \\
p_{\rm g} & = (\hat \gamma -1) (\epsilon_g - n^{'}mc^2). \label{eq:appendix_equation_of_state}
\end{align}

\section{Perturbative approach to the MHD equations: decomposition into fast and slow variables}

Here, we follow similar steps to the mathematical derivation of \cite{LK01}.
The problem at hand naturally possesses two time scales. On the one hand, the rotation period of the pulsar defines the fast time-scale $t_{\rm f} = 2\pi/\Omega$. We further assume that at a distance of a few light-cylinder radii from the central compact object, the pattern of the wind (which consists of the current sheet and the magnetized region) is stationary. The speed of the flow does not change within one pattern, while magnetic field, density, pressure and radiation vary internally (between the hot and cold layers). The system can therefore be thought as an entropy wave comoving with the fluid. This wave evolves on a slow time scale corresponding to the expansion time of the wind $t_{\rm s} \sim 2 \pi r /(r_L \Omega)$. Following \cite{LK01}, the procedure we follow is (1) transform the problem to the slow and fast independent variables, (2) expand the parameters, (3) solve for the zeroth order equations, (4) solve the first order equations requiring that the secular terms they contain vanish.

We first define the phase (over one pattern)
\begin{align}
\Phi = \Omega \left [ t - \int_0^r \frac{dr^{'}}{v_{w}(r^{'})}  \right ],
\end{align}
where $v_w(r)$ is the speed of the pattern, which will be determined later.  Then, the coordinates of the problem are changed from $(r,t)$ to $(R,\Phi)$ where $R= \epsilon r/ r_L$, with $\epsilon \ll 1$ and $R\sim 1$. The Jacobian of the transformation is
\begin{align}
\mathcal{J} = \left ( \begin{tabular}{cc}
$\null \pder[\Phi]{t} = \Omega$ & $\pder[R]{t} = 0$ \\
$\null \pder[\Phi]{r} = -\frac{\Omega}{v_{\omega}}$  & $\pder[R]{r} = \frac{\epsilon}{r_L}$
\end{tabular} \right ).
\end{align}

The continuity equation in the new set of coordinates is
\begin{align}
\pder[\Gamma n^{'}]{\Phi} - \frac{1}{v_w} \pder[\Gamma \beta n^{'}]{\Phi} + \frac{\epsilon}{cR^2} \pder[R^2 \Gamma v n^{'}]{R} = 0,
\end{align}
where the definition of the light-cylinder was used, $r_{\rm L} = c/\Omega$.
The energy equation \ref{eq:appendix_energy} gets the form
\begin{align}
\Omega \pder[T^{00}+ \epsilon_{\rm rad}]{\Phi} + \frac{1}{R^2} \frac{\epsilon}{r_L} \pder[R^2 (T^{01} + F^{r})]{R} - \frac{\Omega}{v_w} \pder[T^{01}+F^{r}]{\Phi} =0, \label{eq:appendix_energy_sf_variable}
\end{align}
where Equation \ref{eq:radiation_energy} was used to express $G^0$. Transforming the entropy Equation \ref{eq:entropy_appendix} is more cumbersome, and after long but simple substitution algebra, it becomes
\begin{align}
& ~~ \left [ \hat \gamma p_{\rm g} + (\hat \gamma -1) h_{\rm rad}^{'}\right ]\pder[\Gamma]{\Phi} + \Gamma \pder[p_{\rm g} + \frac{3}{4} (\hat \gamma-1)h_{\rm rad}^{'} ]{\Phi} - \frac{\hat \gamma p_{\rm g} + (\hat \gamma -1) h_{\rm rad}^{'}}{v_w} \pder[v \Gamma]{\Phi} \nonumber \\
- & ~~ \frac{\Gamma v}{v_w} \pder[p_{\rm g} + \frac{3}{4} (\hat \gamma-1) h_{\rm rad}^{'}]{\Phi} + \frac{\epsilon}{c R^2} \left [ \hat \gamma p_{\rm g} + (\hat \gamma -1)h_{\rm rad}^{'}\right ] \pder[R^2 \Gamma v]{R} \nonumber \\
+ & ~~ \frac{\epsilon \Gamma v}{c} \pder[p_{\rm g} + \frac{3}{4}(\hat \gamma -1 )h_{\rm rad}^{'}]{R} = \frac{\hat \gamma -1}{\Omega} \Gamma (\mathcal{E} - \frac{v}{c} \mathcal{B})j, \label{eq:appendix_entropy_sf_variable}
\end{align}
where the radiation enthalpy $h_{\rm rad}^{'}= 4 \epsilon_{\rm rad}^{'}/3$. 
Finally, the Ampere's and Faraday's equation becomes
\begin{align}
\pder[\mathcal{E}]{\Phi} - \frac{c}{v_w} \pder[\mathcal{B}]{\Phi} + \frac{\epsilon}{R} \pder[R\mathcal{B}]{R} + \frac{4\pi}{\Omega}j &= 0, \\
\pder[\mathcal{B}]{\Phi}- \frac{c}{v_\omega}\pder[\mathcal{E}]{\Phi} + \frac{\epsilon}{R} \pder[R\mathcal{E}]{R}  &= 0. \label{eq:appendix_faraday}
\end{align}

In order to simplify the equations, we introduce $\beta = v/c$ and $\beta_\omega = v_\omega / c $. We further expand all the quantities to the first order in $\epsilon$ such that for a quantity $\mathcal{X}$, we write ${\mathcal{X} = \mathcal{X}_0(\Phi,R)+ \epsilon \mathcal{X}_1(\Phi,R) }$. Since  an entropy wave is considered, we have $\partial \beta_0 / \partial\Phi =0$. 

Expanding the continuity equation in $\epsilon$, the zeroth order gives $\beta_{\omega} = \beta_0(R)$. Therefore, using Ampere and Faraday equations gives
\begin{align}
\pder[\mathcal{B}_0]{\Phi} & = \frac{4\pi \Gamma_0^2\beta_0}{\Omega} j_0. \label{eq:dB_0}
\end{align}
Faraday Equation \ref{eq:appendix_faraday} implies that $\mathcal{E}_0-\beta_0 \mathcal{B}_0$ is independent of $\Phi$. In the region where the assumption of ideal MHD holds, Ohm's law imposes
\begin{align}
\mathcal{E}_0 & = \beta_0 \mathcal{B}_0.
\end{align}
At the zeroth order in $\epsilon$, the energy Equation \ref{eq:appendix_energy_sf_variable} simplifies to
\begin{align}
\pder[p_{\rm g,0} + \frac{\mathcal{B}_{0}^2}{8\pi \Gamma_0^2} +\frac{\epsilon_{\rm rad,0}^{'}}{3} ]{\Phi} =0 \label{eq:appendix_pressure_balance}
\end{align}
which is the pressure balance condition of the current sheet. In addition, it is immediately checked that the entropy equation is satisfied at the zeroth order in $\epsilon$.

In order to describe the internal structure, one needs to consider the first order in $\epsilon$. In this order, the continuity equation becomes
\begin{align}
\frac{\Gamma_0}{\beta_0} \pder[\beta_1 n_0^{'}]{\Phi} = \frac{1}{R^2} \pder[R^2 \Gamma_0 \beta_0 n_0^{'}]{\Phi}, \label{eq:appendix_continuityepsilon}
\end{align}
and the two Maxwell equations
\begin{align}
\pder[\mathcal{E}_1 - \frac{\mathcal{B}_1}{\beta_0}]{\Phi} & = -\frac{1}{R} \pder[R\mathcal{B}_0]{\Phi} - \frac{4\pi}{\Omega} j_1, \\
\pder[\mathcal{B}_1 - \frac{\mathcal{E}_1}{\beta_0}]{\Phi} & = -\frac{1}{R} \pder[R \beta_0 \mathcal{B}_0]{\Phi}. \label{eq:maxB1E1}
\end{align}
Ohm's law reads
\begin{align}
j_1 = \sigma_{\rm c}  [ \Gamma_1 (\mathcal{E}_0 - \beta_0 \mathcal{B}_0) + \Gamma_0 (\mathcal{E}_1 - \beta_0 \mathcal{B}_1) -\Gamma_0 \beta_1 \mathcal{B}_0 ].
\end{align}
The first term is null according to the zeroth order Ohm's law together with the assumption of ideal MHD. One therefore obtains
\begin{align}
\mathcal{E}_1 = \beta_0 \mathcal{B}_1 + \beta_1 \mathcal{B}_0. \label{eq:appendix_idealMHDepsilon}
\end{align} 
Using Equation \ref{eq:appendix_idealMHDepsilon} in the Faraday Equation \ref{eq:maxB1E1} leads to
\begin{align}
\pder[\frac{\beta_1}{\beta_0} \mathcal{B}_0]{\Phi} = \frac{1}{R} \pder[R\beta_0 \mathcal{B}_0]{R}.
\end{align}
Dividing both sides of this last equation by $R^2 \beta_0 \Gamma_0 n_0^{'}$, multiplying both sides of Equation \ref{eq:appendix_continuityepsilon} by $\mathcal{B}_0/(\beta_0 R^3 {n_0^{'}}^2 \Gamma_0^2)$, and taking the difference between them leads to the flux freezing condition after simple algebra
\begin{align}
\pder[\frac{\mathcal{B}_0}{R \Gamma_0 n_0^{'}}]{R} = 0. \label{eq:appendix_fluxfreezing_derivation}
\end{align}
An implicit assumption made in the derivation of this equation is that the density and the magnetic field do not depend on $\Phi$ outside of the current sheet, as is the case for the \textit{striped wind} considered here.

The energy Equation \ref{eq:appendix_energy_sf_variable} becomes
\begin{align}
 \pder[p_{\rm g,1} + (\epsilon_{\rm g,0}-p_{\rm g,1})\Gamma_0^2 \frac{\beta_1}{\beta_0} +\frac{\mathcal{B}_0\mathcal{E}_1}{4\pi \beta_0 \Gamma_0^2} - \epsilon_{\rm rad, 1} + \frac{F_1^{\rm r}}{\beta_0} ]{\Phi} = \frac{1}{R^2} \pder[R^2 \left \{ (\epsilon_{\rm g,0}+p_{\rm g,0}) \Gamma_0^2 \beta_0 +\frac{\mathcal{E}_0 \mathcal{B}_0}{4\pi} +F_0^{\rm r} \right \}  ]{R} \label{eq:app_B18}
\end{align}
Equations \ref{eq:appendix_continuityepsilon} and \ref{eq:app_B18} which represent conservation of particles and energy can be viewed as generalization of Equations A23 to A26 of \cite{LK01}, when radiation terms are included.

The first order entropy Equation \ref{eq:appendix_entropy_sf_variable} can be further simplified by writing the pressure balance condition given by Equation \ref{eq:appendix_pressure_balance} as
\begin{align}
 \pder[p_{\rm g,0}  +\frac{\epsilon_{\rm rad,0}^{'}}{3} ]{\Phi} = - \frac{B_0}{4 \pi \Gamma_0^2}  \pder[B_{0}]{\Phi}.
\end{align}
Using this equation, the entropy Equation \ref{eq:appendix_entropy_sf_variable} in first order of $\epsilon$ obtains the form
\begin{align}
& \frac{\hat{\gamma} p_{\rm g,0} + (\hat{\gamma}-1)\frac{4}{3} \epsilon_{\rm rad,0}^{'}}{R^2} \pder[R^2 \Gamma_0 \beta_0]{R} + \Gamma_0 \beta_0 \pder[p_{\rm g,0} + (\hat{\gamma}-1) \epsilon_{\rm rad,0}^{'}]{R} \nonumber \\
&~~~~~ = \frac{(\hat{\gamma}-1)  (\mathcal{E}_1-\beta_0 \mathcal{B}_1)}{4 \pi \Gamma_0 \beta_0} \pder[\mathcal{B}_0]{\Phi}   + \frac{\Gamma_0}{\beta_0} \pder[\beta_1 (\hat{\gamma}p_{\rm g,0}+\frac{4}{3} (\hat{\gamma}-1)\epsilon_{\rm rad,0}^{'})]{\Phi}
\end{align}
In order to avoid the divergence of the first order terms, integration of the equations over one period of the fast variable should be null. Imposing these regularity conditions is enough to determine the slow variation of all the parameters. The final set of equations reads
\begin{align}
& \pder[R^2 \Gamma_0 \beta_0 \int_0^{2\pi} n_0^{'} d\Phi]{R} = 0 \label{eq:continuityintphi}\\
& \pder[R^2 \Gamma_0^2 \beta_0 \int_0^{2\pi} \left ( (\epsilon_{\rm g,0}+p_{\rm g,0})  +\frac{ \mathcal{B}_0^2}{4\pi \Gamma_0^2} + \frac{F_0^{\rm r}}{\Gamma_0^2 \beta_0} \right )d\Phi ]{R} = 0 \label{eq:energyintphi} \\
& \int_0^{2\pi} \frac{\hat{\gamma} p_{\rm g,0} + (\hat{\gamma}-1)\frac{4}{3}\epsilon_{\rm rad,0}^{'}}{R^2}\pder[R^2 \Gamma_0 \beta_0]{R} d\Phi + \Gamma_0 \beta_0 \pder[\int_0^{2\pi}p_{\rm g,0} + (\hat{\gamma}-1) \epsilon_{\rm rad,0}^{'} d\Phi]{R}  \nonumber \\
& ~~~~~~~~~~~~~ = -\frac{\hat{\gamma}-1}{4\pi\Gamma_0 R} \int_0^{2\pi} \mathcal{B}_0 \pder[R\beta_0 \mathcal{B}_0]{R} d\Phi, \label{eq:entropyintphi}
\end{align}
where the right-hand side of the last equation is obtained by integration by parts and using Maxwell's equations. Note that these equations contain only zeroth order terms and must be satisfied to ensure convergence of higher order terms. Finally, dividing both sides of Equation \ref{eq:entropyintphi} by $\hat \gamma - 1$, leads to Equation \ref{eq:entropyequation}, after recognizing the expression for the total enthalpy (gas and radiation). We note that Equation \ref{eq:energyintphi} can also be simplified by recognizing the expression of the total enthalpy.

\section{Application to the \textit{striped wind} model}

Within the framework of the \textit{striped wind} model, Equations \ref{eq:continuityintphi}, \ref{eq:energyintphi} and \ref{eq:entropyintphi} can be further simplified. Since these equations contain only zeroth order terms, we can omit the subscript 0. Instead, we add subscripts 1 and 2 to refer to the current sheet and the magnetized region respectively. Since the magnetic field changes polarity in between the current sheets, it is best to described the wind using 4 distinct regions such that:
\begin{align}
& \left \{
\begin{aligned}
n^{'} & = n_1^{'}(R) \\
p_{\rm g} & = p_1(R) \\
\mathcal{B} & = 0 \\
\epsilon_{\rm rad}^{'} & = \epsilon_{\rm rad,1}^{'}(R) \\
h^{'} & = h^{'}_{\rm g,1} + h^{'}_{\rm rad,1}
\end{aligned} \right \} \text{for $0<\Phi<\pi \Delta(R)$ and $\pi<\Phi<\pi [1+\Delta(R)]$}
\end{align}
which represents the two current sheets. The two magnetized regions are described by
\begin{align}
& \left \{
\begin{aligned}
n^{'} & = n_2^{'}(R) \\
p_{\rm g} & = p_2(R) \\
\mathcal{B} & = \mathcal{B}(R) \\
\epsilon_{\rm rad}^{'} & = \epsilon_{\rm rad,2}^{'}(R) \\
h^{'} & = h^{'}_{\rm g,2} + h^{'}_{\rm rad,2}
\end{aligned} \right \} \text{for $\pi \Delta(R)<\Phi<\pi $}
\end{align}
\begin{align}
\left \{
\begin{aligned}
n^{'} & = n_2^{'}(R) \\
p_{\rm g} & = p_2(R) \\
\mathcal{B} & = - \mathcal{B}(R) \\
\epsilon_{\rm rad}^{'} & = \epsilon_{\rm rad,2}^{'}(R) \\
h^{'} & = h^{'}_{\rm g,2} + h^{'}_{\rm rad,2}
\end{aligned} \right \} \text{for $\pi[1+ \Delta(R)]<\Phi<2\pi $}
\end{align}

Integration of the continuity Equation \ref{eq:continuityintphi} over $\Phi$ can be directly performed using $\pder[\beta]{\Phi} = 0$,
\begin{align}
\pder[R^2 \Gamma \beta \left \{ (1-\Delta) n_2^{'} + \Delta n_1^{'} \right \}]{R} =0, \label{eq:apcontC4}
\end{align}
which is Equation \ref{eq:continuityderivative} in the main text. Similarly, the energy Equation \ref{eq:energyintphi} becomes
\begin{align}
\pder[R^2 \Gamma^2 \beta \left \{ (1-\Delta) \left ( h_2^{'} + \frac{{\mathcal{B}^{'}}^2}{4\pi} \right ) + \Delta h_1^{'} \right \}]{R} =0, \label{eq:apenerC5}
\end{align}
and the entropy Equation \ref{eq:entropyintphi} is immediately reduced to Equation \ref{eq:entropyequation}.

In order to obtain the numerical solution, we solved the equations of continuity \ref{eq:apcontC4}, energy \ref{eq:apenerC5} and entropy \ref{eq:entropyintphi}. These are combined with the flux-freezing condition $\pder[\frac{{\mathcal{B}^{'}}}{Rn_2^{'}}]{R} = 0$, as well as the assumption of steady reconnection rate (Equation 9 in the main text),
\begin{align}
\pder[\beta \Gamma R {\mathcal{B}}^{'}]{R} = -A \frac{R \mathcal{B}^{'}}{ \Gamma}. \nonumber
\end{align}
In case I, the energy and entropy equations obtain the form
\begin{align}
& \pder[R^2 \beta \Gamma^2 \left ( m_p c^2 ( n_1^{'} \Delta + (1-\Delta) n_2^{'} ) + \frac{1+\Delta}{4\pi} {\mathcal{B}^{'}}^2  \right ) ]{R } = 0, \\
& \frac{4}{R^2}  \Delta {\mathcal{B}^{'}}^2 \pder[R^2 \Gamma \beta]{R} +  3 \Gamma \beta \pder[ \Delta {\mathcal{B}^{'}}^2 ]{R} +  \frac{\beta}{\Gamma} \pder[\Gamma^2 (1-\Delta) {\mathcal{B}^{'}}^2]{R} +\nonumber\\
 & ~~~~~~~~~~~~ +\frac{2\Gamma (1-\Delta) {\mathcal{B}^{'}}^2 }{R} \pder[\beta R]{R} = 0, 
\end{align}
where we made use of the pressure balance condition, $p_1 = {\mathcal{B}^{'}}^{2}/(8\pi)$.
For case II, $\hat \gamma =5/3 $ and the energy and entropy equations become
\begin{align}
& \pder[R^2 \Gamma^2 \beta \left \{  m_p c^2 ( \Delta n_1^{'} + (1- \Delta) n_2^{'} ) + \frac{5}{2} k_{\rm B} n_2^{'} T^{'}  + \frac{4}{3} a_{\rm th} {T^{'}}^4 +   \left (2+\frac{1}{2}\Delta \right ) \frac{{\mathcal{B}^{'}}^2}{8\pi}  \right \} ]{R} = 0, \\
& \frac{4\pi}{R^2} \left [ 5 k_{\rm B} n_2^{'} T^{'} + \frac{8}{3}  a_{\rm th} {T^{'}}^4 + \frac{5}{2} \Delta \frac{{\mathcal{B}^{'}}^2}{4 \pi}  \right ]  \pder[R^2 \Gamma \beta]{R} +3\pi \Gamma \beta \pder[4 k_{\rm B} n_2^{'} T^{'} + \frac{16}{3} a_{\rm th} {T^{'}}^4 + \Delta \frac{{\mathcal{B}^{'}}^2}{\pi }]{R}+ \nonumber \\ 
 & ~~~~~~~~~~~~ + \frac{\beta}{\Gamma} \pder[\Gamma^2 (1-\Delta) {\mathcal{B}^{'}}^2]{R} +\frac{2\Gamma (1-\Delta) {\mathcal{B}^{'}}^2 }{R} \pder[\beta R]{R} = 0.
\end{align}
In addition in this case, pressure balance condition gives
\begin{align}
 T^{'} = \frac{{\mathcal{B}^{'}}^2 }{8\pi k_B (n_1^{'}-n_2^{'})}. \label{eq:appendix_caseII_total2}
\end{align}

\section{Expansion of MHD equations to first order in $\Delta$, $\sigma^{-1}$ and $\Gamma^{-1}$}

The assumptions $\Delta \ll 1$ , $\sigma \equiv \mathcal{B^{'}}^2/(4\pi n^{'} m_p c ^2) \gg 1$ and $\Gamma \gg 1$ lead to a degeneracy. Therefore, in order to obtain analytical solutions to the set of MHD equations, one needs to consider their first order expansion in $\Delta$, $\sigma^{-1}$ and $\Gamma^{-1}$. This can be directly seen since in the zeroth approximation with respect to the small parameters $\Delta$, $\sigma^{-1}$ and $\Gamma^{-1}$ 
the continuity Equation \ref{eq:continuityderivative}, the energy Equation \ref{eq:energyderivative} and the flux freezing condition Equation \ref{eq:fluxfreezingderivative} become
\begin{align}
R^2 \Gamma  n_2^{'} & =  R^2_0 \Gamma_0  n^{'}_0, \label{eq:D1}\\
  \Gamma^2 R^2 \mathcal{B^{'}}^2 & =  \Gamma_0^2 R_0^2 {\mathcal{B}_0^{'}}^2, \label{eq:D2} \\
\frac{\mathcal{B^{'}} }{ R n_2^{'} } & = \frac{\mathcal{B}_0^{'} }{ R_0 n_0^{'} }. \label{eq:D3}
\end{align}
This set of equations is degenerated because if one divide Equation \ref{eq:D2} by Equation \ref{eq:D3} squared, one gets Equation \ref{eq:D1} squared.

First order expansion in $\Delta$ and $\sigma$ of the continuity Equation \ref{eq:continuityequationexpanded} leads to
\begin{align}
\beta \Gamma R^2 n_2^{'} - \beta_0 \Gamma_0 R_0^2 n_0^{'} =\beta \Gamma \Delta R^2 n_2^{'} \left (1-\frac{n_1^{'}}{n_2^{'}}\right ).
\end{align}
Because the right-hand side of the equation is of order $\mathcal{O}(\Delta)$, the zeroth order of the continuity equation, given by $\Gamma R^2 n_2^{'} = \Gamma_0 R_0^2 n_0^{'}$, can safely be used to give
\begin{align}
\beta \Gamma R^2 n_2^{'} - \beta_0 \Gamma_0 R_0^2 n_0^{'} =\beta_0 \Gamma_0 \Delta R_0^2 n_0^{'} \left (1-\frac{n_1^{'}}{n_2^{'}}\right ). \label{eq:appendix_continuity1order}
\end{align}
Using Equation \ref{eq:fluxfreezingexpanded} from the main text, the energy Equation \ref{eq:energyequationexpanded} is expended to get
\begin{align}
\frac{\mathcal{B^{'}}^2_0}{4\pi} \frac{ \beta \Gamma^2 R^4 {n_2^{'}}^{2} - \beta_0 \Gamma_0^2 R_0^4 {n_0^{'}}^{2}}{R_0^2 n_0^2} = c \left [ \Delta \left ( h_2^{'} - h_1^{'} + \frac{\mathcal{B^{'}}^2}{4\pi} \right ) - h_2^{'}\right ] \Gamma^2 R^2 + m_p c^3 n_0^{'} \Gamma_0^2 R_0^2. \label{eq:appendix_energy1order}
\end{align}
Using Equation \ref{eq:appendix_continuity1order}, the expression on the left-hand side of Equation \ref{eq:appendix_energy1order} may be expressed via small quantities
\begin{align}
\frac{\beta \Gamma^2 R^4 {n_2^{'}}^2-\beta_0 \Gamma_0^2 R_0^4 {n_0^{'}}^2}{R_0^2 {n_0^{'}}^2} & = \frac{\beta^2 \Gamma^2 R^4 {n_2^{'}}^2-\beta_0^2 \Gamma_0^2 R_0^4 {n_0^{'}}^2}{c R_0^2 {n_0^{'}}^2} + \frac{c R^4 {n_2^{'}}^2}{2R_0^2 {n_0^{'}}^2}-\frac{c R_0^2}{2} \nonumber \\
& = \frac{2 \Gamma_0 (\beta \Gamma R^2 {n_2^{'}} - \beta_0 \Gamma_0 R_0^2 n_0^{'})}{n_0^{'}} + \frac{c R^2_0}{2} \left ( \frac{\Gamma_0^2}{\Gamma^2} -1  \right ) \nonumber \\
& = c R_0^2 \Gamma_0^2 \left [ 2\Delta \left ( 1- \frac{n_1^{'}}{n_2^{'}} \right ) + \frac{1}{2\Gamma^2} - \frac{1}{2 \Gamma_0^2}\right ].
\end{align}
Therefore, the energy Equation \ref{eq:appendix_energy1order} is written in a form containing only small quantities
\begin{align}
2\Delta \left ( 1-\frac{n_1^{'}}{n_2^{'}}\right ) + \left [ h_2^{'}  - \Delta \left( h_2^{'}-h_1^{'} + \frac{\mathcal{B^{'}}^2}{4\pi}\right )\right ] \frac{\Gamma^2 R^2 }{\Gamma_0^2 R_0^2 } \frac{4\pi}{\mathcal{B^{'}}_0^2} + \frac{1}{2\Gamma^2} = \frac{1}{\sigma_0} + \frac{1}{2\Gamma_0^2}. \label{eq:D57}
\end{align}
Using the zeroth order continuity Equation, $\Gamma R^2 n_2^{'} = \Gamma_0 R_0^2 n_0^{'}$ and the flux freezing condition (Equation \ref{eq:fluxfreezingexpanded}), one can write
\begin{align}
\frac{\Gamma^2 R^2}{\Gamma_0^2 R_0^2} \frac{4\pi}{\mathcal{B^{'}}_0^2} = \frac{\Gamma}{\Gamma_{\rm max} m_p c^2 n_2^{'}}, \label{eq:appendix_bzerogamma2}
\end{align}
where $\Gamma_{\rm max} = \sigma_0 \Gamma_0 $ is the maximum achievable Lorentz factor.

The reconnection layer is in hydrostatic equilibrium,
\begin{align}
p_1 = p_2 + \frac{\mathcal{B^{'}}^2}{8\pi}
\end{align}
where $p_1$ and $p_2$ are the total (gas and radiation) pressure respectively in the current sheet and in the magnetized part. One therefore finds
\begin{align}
h_1^{'}-h_2^{'}-\frac{\mathcal{B}'^2}{4\pi}=\epsilon_1^{'}-\epsilon_2^{'}-\frac{\mathcal{B}'^2}{8\pi}.
\end{align}
Substituting this, as well as Equation \ref{eq:appendix_bzerogamma2}, into Equation \ref{eq:D57}  yields
\begin{align}
\Delta\left(  2-2\frac{n'_1}{n'_2}+\frac{(\epsilon_1^{'}-\epsilon_2^{'})\Gamma}{\Gamma_{\rm max}m_pc^2n'_2}-
 \frac{1}{2}\frac{\Gamma^2R^2\mathcal{B}'^2}{\Gamma^2_0R^2_0\mathcal{B}_0'^2}\right)+
\frac{h_2^{'} \Gamma}{\Gamma_{\rm max}m_pc^2n'_2}=\frac 1{\sigma_0}+\frac 1{2\Gamma_0^2}-\frac 1{2\Gamma^2}
.\end{align}
 It follows from Equation \ref{eq:D2} that the last term in the brackets is unity. Moreover, beyond the fast magnetosonic point, one can neglect the terms in the right-hand side as compared with the second term in the left-hand side. Therefore the energy equation obtains the final form
\begin{align}
\Delta\left(2\frac{n'_1}{n'_2}-\frac { 3}{2}-\frac{(\epsilon_1^{'}-\epsilon_2^{'})\Gamma}{\Gamma_{\rm max}m_pc^2n'_2}\right)=
\frac{h_2^{'}\Gamma}{\Gamma_{\rm max}m_pc^2n'_2} \label{eq:D62}
\end{align}

The equation describing the reconnection rate can be written to first order in $\Delta$ as follows. We first use the flux-freezing condition (Equation \ref{eq:fluxfreezingexpanded}) to write Equation \ref{eq:reconnectionrate} as
\begin{align}
\pder[\beta \Gamma R^2 {n_2^{'}}]{R} = -A \frac{R^2 n_2^{'}}{\Gamma}. \label{eq:appendix_25}
\end{align}
The left-hand side vanishes when using the zeroth order continuity equation. Therefore, one uses the first order continuity equation in the left-hand side and the zeroth order continuity equation in the right-hand side. The reconnection rate Equation \ref{eq:appendix_25} becomes
\begin{align}
\pder[\Delta \left( \frac{n_1^{'}}{n_2^{'}} -1\right )]{R} = \frac{A}{\Gamma^2}. \label{eq:appendix_reconnectionratefinal}
\end{align}

The entropy Equation \ref{eq:entropyequation}, is written as 
\begin{align}
\langle p \rangle \pder[\beta \Gamma R^2 ]{R} + \pder[\langle \epsilon^{'} \rangle \beta \Gamma R^2]{R} = -\frac{R}{4\pi} \big \langle \mathcal{B^{'}} \pder[\beta \Gamma R \mathcal{B^{'}}]{R} \big \rangle. \label{eq:appendix_entropyexpanded}
\end{align}
The averaging in the right-hand side should be performed with care due to discontinuity of the magnetic field at the boundary of the current sheet:
\begin{align}
\big \langle \mathcal{B^{'}} \pder[\beta \Gamma R \mathcal{B^{'}}]{R} \big \rangle & =  \langle \mathcal{B^{'}}^2 \rangle \pder[\beta \Gamma R]{R} + \frac{\beta \Gamma R}{2}  \pder[\langle \mathcal{B^{'}}^2 \rangle]{R} \nonumber \\
& =  (1-\Delta) \mathcal{B^{'}}^2 \pder[\beta \Gamma R]{R} + \frac{\beta \Gamma R}{2} \pder[(1-\Delta) \mathcal{B^{'}}^2]{R}    \nonumber \\
& =  \mathcal{B^{'}} \pder[\beta \Gamma R \mathcal{B^{'}}]{R} - \Delta \mathcal{B^{'}} \pder[R\Gamma \mathcal{B^{'}}]{R} - \frac{1}{2} R \Gamma \mathcal{B}^2 \pder[\Delta]{R} . \label{eq:appendix_trick_magnetic}
\end{align}
This expression is further simplified by using the flux freezing condition (Equation \ref{eq:fluxfreezingexpanded}) and the first order continuity Equation \ref{eq:appendix_continuity1order}. After some algebra, one gets
\begin{align}
\big \langle \mathcal{B^{'}} \pder[\beta \Gamma R \mathcal{B^{'}}]{R} \big \rangle = 4\pi \Gamma_{\rm max} m_p c^3 n_2^{'} R \pder[\Delta \left ( \frac{1}{2} - \frac{n_1^{'}}{n_2^{'}} \right )]{R}.
\end{align}
Using this results in the entropy equation (\ref{eq:appendix_entropyexpanded}) gives its final expression,
\begin{align}
\frac{p_1 \Delta + (1-\Delta)p_2}{m_p c^2 n_2^{'} R^2 } \pder[\beta \Gamma R^2]{R} + \frac{1}{m_p c^2 n_2^{'} R^2 } \pder[\beta \Gamma R^2 (\epsilon_1^{'} \Delta + (1-\Delta) \epsilon_2^{'}]{R} \nonumber  \\ = c \Gamma_{\rm max} \pder [\Delta \left ( \frac{n_1^{'}}{n_2^{'}} -\frac{1}{2}  \right )]{R}. ~~~~~~~~~~~~~~ \label{eq:appendix_heatbalacecondition}
\end{align}

\section{Analytical solutions to the MHD equations in the first order approximation}

\subsection{Case I: the heat remains in the reconnection layer}

If the heat released by magnetic reconnection remains in the current sheet, then the magnetized region is cold and one has $p_2 = 0$ and $\epsilon^{'}_2 = m_p c^2 n_2^{'}$. It also implies that in the current sheet, the thermal energy and the pressure are dominated by the radiation so that $\epsilon_1^{'} = 3p_1 + m_p c^2 n_1^{'}$. The pressure balance condition becomes 
\begin{align}
p_1 = \frac{ \mathcal{B^{'}}^2}{8\pi}   = \frac{\sigma_0 R^2 {n_2^{'}}^2 m_p c^2}{2 R_0^2 n_0^{'}}. \label{eq:app_E1}
\end{align}
In this case,
\begin{align}
\epsilon_1-\epsilon_2=3p_1+m_pc^2(n'_1-n'_2)=\frac 32\frac{\sigma_0R^2n'^2_2m_pc^2}{R_0^2n'_0}+m_pc^2(n'_1-n'_2).
\end{align}
In addition, the heat balance Equation \ref{eq:heatbalacecondition} can be simplified to
\begin{align}
\frac{2\sigma_0 \Delta n_2^{'}}{R_0^2 n_0^{'}} \pder[\Gamma R^2]{R} + \frac{\Gamma}{n_2^{'}} \pder[\frac{3}{2} \Delta \sigma_0 \frac{R^2 {n_2^{'}}^2}{R_0^2 n_0^{'}} ]{R} = \Gamma_{\rm max} \pder[ \Delta \left ( \frac{n_1^{'}}{n_2^{'}} - \frac{1}{2} \right ) ]{R}, \label{eq:heatbalance1stintermediaire}
\end{align}

The system of equations to be solved is composed of the energy Equation \ref{eq:energyfluxfinal}, the reconnection rate Equation \ref{eq:reconnectionratefinal} and the heat balance condition Equation \ref{eq:heatbalance1stintermediaire}. The system can be simplified to
\begin{align}
\frac{\Gamma}{\Gamma_{\rm max}} & = \Delta \left ( 2 \frac{n_1^{'}}{n_2^{'}} -3\right ), \label{eq:app_E3}\\
\pder[\frac{\Gamma}{ \Gamma_{\rm max}} + \Delta ]{R} & = \frac{2A}{\Gamma^2}, \\
2 \pder[\Delta]{R} - \frac{\Delta}{\Gamma} \pder[\Gamma]{R} + \frac{\Delta}{R} & = \pder[\frac{n_1^{'} \Delta}{n_2^{'}}]{R},
\end{align}
where the last equation was obtained by using the zeroth order continuity Equation in Equation \ref{eq:heatbalance1stintermediaire} to eliminate $n_2^{'}$. These equations are satisfied by the ansatz $\gamma \propto \Delta \propto R^{1/3}$. Substituting and looking for the coefficients, we obtain Equations \ref{eq:lorentz_factr_case1}, \ref{eq:delta_case1} and \ref{eq:ratiodensitiescase1}.

\subsection{Case II: the heat is redistributed in the magnetized region}

In case II, the heat is assumed to be redistributed in the magnetized region by radiation, which fills the striped wind with energy density $\epsilon_{\rm rad}^{'}$ and the temperature both inside and outside the sheets remains non-relativistic  $k_{\rm B} T^{'} \ll m_e c^2$. As a result, one can write $\epsilon_{1,2}^{'} = \epsilon^{'}_{\rm rad} + m_p c^2 n_{1,2}^{'}$, $p_1 = n_1^{'} k_{\rm B} T^{'} + (1/3)\epsilon_{\rm rad}^{'}$ and $p_2 = (1/3) \epsilon_{\rm rad}^{'}$. Therefore, the heat balance Equation \ref{eq:heatbalacecondition} can be written as
\begin{align}
\frac{1}{R^2} \left ( \frac{\epsilon_{\rm rad}^{'}}{3m_p c^2 n_2^{'}} + \Delta \frac{n_1^{'}}{n_2^{'}} \frac{k_{\rm B} T^{'}}{m_p c} \right ) \pder[\Gamma R^2]{R} + \frac{1}{m_p c^2 n_2^{'} R^2} \pder[\epsilon_{\rm rad}^{'} \Gamma R^2]{R} = \Gamma_{\rm max} \pder[\Delta \left ( \frac{n_1^{'}}{n_2^{'}} - \frac{1}{2} \right )]{R}. \label{eq:entropyfluxcase2intermediate}
\end{align}
The density in the current sheet should be very large to balance the magnetic pressure, and therefore $n_1^{'} \gg n_2^{'}$.  Therefore, one can neglect the second term in the brackets of the left hand side of the energy Equation \ref{eq:D62}. The last term in the brackets is reduced to
\begin{align}
\frac{(\epsilon_1^{'}-\epsilon_2^{'})\Gamma}{\Gamma_{\rm max}m_pc^2n'_2}=\left(\frac{n'_1}{n'_2}-1\right)
\frac{\Gamma}{\Gamma_{\rm max}}\approx \frac{n'_1}{n'_2}
\frac{\Gamma}{\Gamma_{\rm max}}.
\end{align}
In the asymptotic region $\Gamma_0\ll\Gamma\ll\Gamma_{\rm max}$ this term could also be neglected. Under these conditions, the energy flux Equation \ref{eq:energyfluxfinal}, the reconnection rate Equation \ref{eq:reconnectionratefinal}, and the heat balance condition Equation \ref{eq:entropyfluxcase2intermediate} are simplified to
\begin{align}
\left ( 1+ \frac{4 \epsilon_{\rm rad}^{'}}{3 m_p c^2 n_2^{'}} \right )\frac{\Gamma }{\Gamma_{\rm max}} & = \frac{2 n_1^{'} \Delta}{n_2^{'}}, \\
\pder[\frac{n_1^{'} \Delta}{n_2^{'}}]{R} & = \frac{A}{\Gamma^2}, \\
\pder[\frac{\epsilon_{\rm rad}^{'}}{m_p n_2^{'}}]{R} + \frac{1}{3} \frac{\epsilon_{\rm rad}^{'}}{m_p n_2^{'}} \frac{1}{ \Gamma R} \pder[\Gamma R^2]{R} & = \frac{\Gamma_{\rm max}}{\Gamma} \pder[\frac{n_1^{'} \Delta}{n_2^{'}}]{R}.
\end{align}
These equations have the solution given by Equations \ref{eq:30}, \ref{eq:ratiodeltan1n2} and \ref{eq:temperaturecase2}.

\bibliographystyle{mnras}
\bibliography{/home/cayley/Desktop/Thesis/Publication/AAAA____Bibliography/biblio}


\end{document}